\documentclass[usegraphicx,usenatbib]{mn2e}
\usepackage{times,amssymb,amsbsy}

\newcounter{eqold}
\newcounter{eqbid}
\newcommand{\D}[2]{\frac{{\rm d} #2}{{\rm d} #1}}
\newcommand{\DD}[2]{\frac{{\rm d}^2 #2}{{\rm d} #1^2}}
\newcommand\bb[1]{\mbox{\boldmath{$#1$}}}
\newcommand\grad{\bb{\nabla}}
\newcommand\bcdot{\bb{\cdot}}
\newcommand\btimes{\bb{\times}}
\newcommand{\mc}[1]{\mathcal{#1}}

\newcommand{\msb}[1]{\bb{\mathsf{#1}}}

\newcommand{\eb}{\hat{\bb{b}}}
\newcommand{\ex}{\hat{\bb{x}}}
\newcommand{\ey}{\hat{\bb{y}}}
\newcommand{\ez}{\hat{\bb{z}}}
\newcommand{\vasq}{v^2_{\rm A}}
\newcommand{\vthsq}{v^2_{\rm th}}
\newcommand{\mfp}{\lambda_{\rm mfp}}
\newcommand{\cond}{\omega_{\rm cond}}
\newcommand{\visc}{\omega_{\rm visc}}
\newcommand{\dyn}{\omega_{\rm dyn}}

\title[Dynamical stability of the ICM]
{Dynamical stability of a thermally stratified intracluster medium with anisotropic momentum and heat transport}
\author[M. W. Kunz]
{Matthew W. Kunz\thanks{E-mail: kunz@thphys.ox.ac.uk} \\
Rudolf Peierls Centre for Theoretical Physics, University of Oxford, 1 Keble Road, Oxford OX1 3NP, U. K.}
\date{Submitted 2011 April 18}
\pagerange{\pageref{firstpage}--\pageref{lastpage}} \pubyear{2011}
\def\LaTeX{L\kern-.36em\raise.3ex\hbox{a}\kern-.15em
    T\kern-.1667em\lower.7ex\hbox{E}\kern-.125emX}
\begin{document}
\label{firstpage} \maketitle

\begin{abstract}
In weakly-collisional plasmas such as the intracluster medium (ICM), heat and momentum transport become anisotropic with respect to the local magnetic field direction. Anisotropic heat conduction causes the slow magnetosonic wave to become buoyantly unstable to the magnetothermal instability (MTI) when the temperature increases in the direction of gravity and to the heat-flux--driven buoyancy instability (HBI) when the temperature decreases in the direction of gravity. The local changes in magnetic field strength that attend these instabilities cause pressure anisotropies that viscously damp motions parallel to the magnetic field. In this paper we employ a linear stability analysis to elucidate the effects of anisotropic viscosity (i.e. Braginskii pressure anisotropy) on the MTI and HBI. By stifling the convergence/divergence of magnetic field lines, pressure anisotropy significantly affects how the ICM interacts with the temperature gradient. Instabilities which depend upon the convergence/divergence of magnetic field lines to generate unstable buoyant motions (the HBI) are suppressed over much of the wavenumber space, whereas those which are otherwise impeded by field-line convergence/divergence (the MTI) are strengthened. As a result, the wavenumbers at which the HBI survives largely unsuppressed in the ICM have parallel components too small to rigorously be considered local. This is particularly true as the magnetic field becomes more and more orthogonal to the temperature gradient. The field-line insulation found by recent numerical simulations to be a nonlinear consequence of the standard HBI might therefore be attenuated. In contrast, the fastest-growing MTI modes are unaffected by anisotropic viscosity. However, we find that anisotropic viscosity couples slow and Alfv\'{e}n waves in such a way as to buoyantly destabilise Alfv\'{e}nic fluctuations when the temperature increases in the direction of gravity. Consequently, many wavenumbers previously considered MTI-stable or slow-growing are in fact maximally unstable. We discuss the physical interpretation of these instabilities in detail.
\end{abstract}


\section{Introduction}

The intracluster medium (ICM) is stratified, not only in pressure, but also in entropy. Until recently, the latter was thought to be of paramount importance to the ICM's dynamical stability. The reason for this is easy to understand. In an atmosphere where entropy increases upwards, an upward adiabatic displacement of a fluid element leaves the element cooler than its surroundings. A cool element is denser (because of local pressure balance), so the buoyancy force is restoring. If, on the other hand, the entropy were to decrease upwards, an upward adiabatic displacement produces a fluid element that is warmer than its surroundings and so there is no restoring buoyancy force, the fluid element continues to rise, and convective instability ensues \citep{schwarzschild58}. Careful observations have shown that the ICM generically has a positive entropy gradient \citep[e.g.][]{cdvs09} and so, by this reasoning, is convectively stable.

In fact, matters are not this simple. The ICM is not a conventional fluid. Firstly, it is magnetized. Secondly, particle-particle collisions are rare. The conductive flow of heat consequently becomes strongly anisotropic with respect to the local magnetic field direction, since collisional energy exchange by (predominantly) the electrons can occur much more readily along the magnetic field than across it. The heat is then restricted to being channeled along magnetic lines of force. When the conduction timescale is shorter than any other dynamical timescale in the system, magnetic field lines become isotherms.

Such a radical change in the thermal behaviour of a weakly collisional plasma profoundly alters its stability properties. Drawing on the powerful analogy between angular momentum and entropy stratification, \citet{balbus00} argued that the temperature gradient takes precedence over the entropy gradient in determining the convective stability of a weakly collisional plasma. He supported this conjecture with a linear stability analysis that proved the existence of the magnetothermal instability \citep[hereafter, MTI;][]{balbus00,balbus01}, which is triggered in regions where the temperature (as opposed to the entropy) increases in the direction of gravity. Subsequent numerical work by \citet{ps05,ps07} and \citet{mpsq10} demonstrated the efficacy of the MTI and extended it into the non-linear regime. Numerical studies of the effect of the MTI on the evolution of the outer regions of non-isothermal galaxy clusters, where the temperature decreases with distance from the central core, followed soon thereafter \citep{psl08}.

In contrast, the inner $\sim$$200~{\rm kpc}$ or so of non-isothermal clusters are characterised by outwardly increasing temperature profiles \citep[e.g.][]{pjkt05,vmmjfs05}. \citet{quataert08} showed that the ICM in these inverted temperature profiles is also buoyantly unstable to a heat-flux--driven instability, now referred to as the HBI. The HBI arises because perturbed fluid elements are heated/cooled by a background heat flux in such a way as to become buoyantly unstable. Recently there has been a surge of numerical efforts to understand the non-linear evolution of the HBI and its implications for the so-called `cooling-flow problem' exhibited by cool-core clusters \citep{pq08,pqs09,brbp09,pqs10,ro10,mpsq10,mty11}.

In this paper, we extend the work of \citet{balbus00} and \citet{quataert08} to include the effects of pressure anisotropy (i.e. anisotropic viscosity). We are motivated by the simple consideration that one cannot self-consistently take the limit of fast thermal conduction along magnetic field lines while simultaneously neglecting differences between the thermal pressure parallel and perpendicular to the magnetic field direction. While the dynamical effects of pressure anisotropy occur on a timescale longer than conduction, in weakly collisional plasmas such as the ICM this timescale is nevertheless still shorter than (or at least as short as) the dynamical timescale and, as a consequence, the MTI and HBI growth times. Our principal result is that, by stifling the convergence/divergence of magnetic field lines, pressure anisotropy significantly affects how the plasma in the ICM interacts with the temperature gradient. Instabilities which depend upon the convergence/divergence of magnetic field lines to generate unstable buoyant motions (the HBI) are suppressed over much of the wavenumber space, whereas those which are otherwise impeded by field-line convergence/divergence (the MTI) are strengthened. We also comment on how pressure anisotropy affects the heat-flux--driven buoyancy overstability recently found by \citet{br10}.

The paper is organised as follows. In Section \ref{sec:motivation}, we motivate the inclusion of pressure anisotropy in our picture of weakly collisional buoyancy instabilities and provide a qualitative discussion of its effects. Readers not interested in the mathematical details may read this Section and proceed immediately to the conclusions (\S\ref{sec:discussion}). In Section \ref{sec:formulation}, we formulate the problem by presenting the basic equations, linearising them, and deriving the dispersion relation governing small perturbations about a simple equilibrium state. The solutions of this dispersion relation are examined in Section \ref{sec:results}. We conclude in Section \ref{sec:discussion} with a summary of our results and a brief discussion of their implications for the structure and evolution of the ICM.

\section{Physical Motivation and Theoretical Expectations}\label{sec:motivation}

The HBI and MTI operate most efficiently at sufficiently small wavelengths for which the conduction rate is much greater than the local dynamical frequency, i.e. $\cond \gg \dyn \equiv ( g / H )^{1/2} = v_{\rm th} / H$, where $g$ is the gravitational acceleration, $H$ is the thermal-pressure scale-height of the plasma, and $v_{\rm th}$ is the thermal speed of the ions. This precludes the usual buoyant restoring force, which would otherwise result in Brunt-V\"{a}is\"{a}l\"{a} oscillations, by ensuring magnetically-tethered fluid elements communicate thermodynamically much faster with one another than they do with the ambient medium. In weakly collisional environments such as the ICM, this timescale separation is satisfied for a wide range of wavenumbers satisfying $k_{||} (\mfp H)^{1/2} \gg (m_{\rm e} / m_{\rm i})^{1/4} \sim 0.1$, where $k_{||}$ is the wavenumber along the magnetic field, $\mfp = v_{\rm th} / \nu_{\rm i}$ is the particle mean free path between collisions, and $\nu_{\rm i}$ is the ion-ion collision frequency. Typical values of $H / \mfp$ in the ICM decrease outwards from $\sim$$10^3$ to $\sim$$10^2$ in the cool-cores of non-isothermal clusters, and from $\sim$$10^2$ to $\sim$$10$ beyond the cooling radius out to $\sim$$1~{\rm Mpc}$. It is important to note that, while the HBI and MTI owe their existence to rapid conduction along magnetic field lines, the unstable perturbations themselves only grow at a rate $\sim$$\dyn$. This is because the free energy required to drive the instabilities is extracted from the background temperature gradient, which is set by macroscale processes, at a rate determined by gravity.

There is however another timescale that ought to be considered. In a magnetized plasma, any change in magnetic field strength must be accompanied by a corresponding change in the perpendicular gas pressure, since the first adiabatic invariant for each particle is conserved on timescales much longer than the inverse of the ion cyclotron frequency (which is extremely short in the ICM; see \S2.2 of \citealt{kscbs11}). The resulting pressure anisotropy is the physical effect behind what is known as \citet{braginskii65} viscosity -- the restriction of the viscous damping (to dominant order in the Larmor radius expansion) to the motions and gradients parallel to the magnetic field.\footnote{Pressure anisotropy also leads to microscale plasma instabilities \citep[e.g. see][and references therein]{sckhs05} but here we will consider perturbations around equilibria that do not trigger those.} This in turn implies that perturbations in magnetic field strength are erased at the viscous damping rate $\visc$ (see eq. \ref{eqn:wvisc}). While $\visc$ is smaller than $\cond$ by a factor $\sim$$10$ (see end of Section \ref{sec:lineareqns}), and so it may be tempting to ignore viscosity relative to conduction, $\visc$ is in fact much greater than the growth rates of the MTI and HBI at most wavelengths at which the latter are usually thought to operate in the ICM. Here we provide a qualitative discussion of these effects.

\begin{figure}
\centering
\includegraphics[width=2.5in]{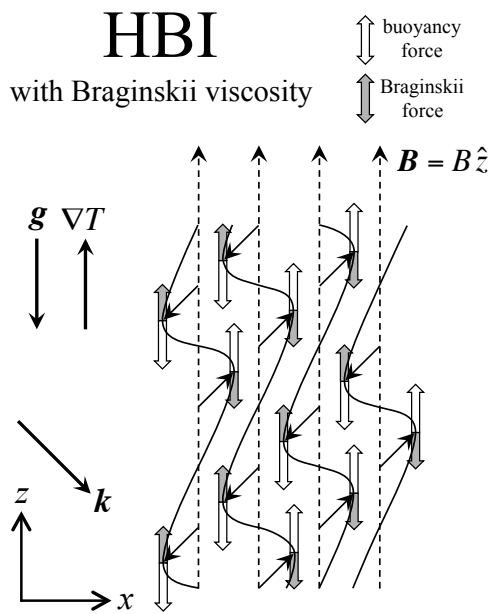}
\caption{HBI subject to Braginskii viscosity. The plasma is threaded by a vertical magnetic field (dashed lines) and has a background heat flux in the $-z$ direction. A perturbation (black arrows) with non-zero $k_x$ and $k_z$ modifies the field lines as illustrated (black curves). The heat flux, forced to follow the perturbed field lines, converges and diverges, leading to heating and cooling of the plasma. For a plasma with ${\rm d}T/{\rm d}z>0$, a downwardly-displaced fluid element loses energy, causing it to sink deeper in the gravitational field (and vice-versa for an upwardly-displaced fluid element). The buoyancy force responsible for this behaviour is denoted by the white solid arrows. The pressure anisotropy, which is generated by motions along the background field lines, contributes a Braginskii (viscous) force (denoted by the grey solid arrows) that impedes this motion. For wavelengths such that $\visc \gg \dyn$, the two forces become nearly equal and opposite and the convergence (divergence) of field lines responsible for the HBI is wiped away faster than upwardly (downwardly) displaced fluid elements can take advantage of the increased (decreased) heating.}
\label{fig:hbi}
\end{figure}

The HBI relies on the presence of a background heat flux, which may be tapped into by the convergence and divergence of conducting magnetic field lines. Downwardly-displaced fluid elements find themselves in regions where field lines diverge; they are conductively cooled via the background heat flux, lose energy, and sink further down in the gravitational potential. As they sink, the local field lines diverge further and an instability ensues. By contrast, an upwardly-displaced fluid element gains energy from the converging heat flux and thus buoyantly rises. Braginskii viscosity hinders the HBI by damping perturbations to the magnetic field strength and thereby preventing convergence and divergence of field lines (see Fig. \ref{fig:hbi}). If $\visc \gg \dyn$, the convergence (divergence) of field lines responsible for the HBI is wiped away faster than upwardly (downwardly) displaced fluid elements can take advantage of the increased (decreased) heating. In fact, as we show in Section \ref{sec:hbi}, the buoyancy and viscous forces become nearly equal and opposite when the background field is vertical and $\visc \gg \dyn$. If the fluid element were to rise buoyantly, it would locally increase the magnetic field strength and generate a pressure anisotropy, which would cause a viscous stress that damps the vertical motion and halts the HBI. Pressure anisotropy can be therefore thought of as providing an effective tension that `tethers' a buoyant fluid element to its original location, preventing it from rising. The only HBI modes to evade strong suppression are those which have wavenumbers satisfying $\visc \lesssim \dyn \lesssim \cond$; these modes are not the same as those usually thought to be the fastest growing.

%
%

%
%
\begin{figure}
\centering
\includegraphics[angle=90,width=3.2in]{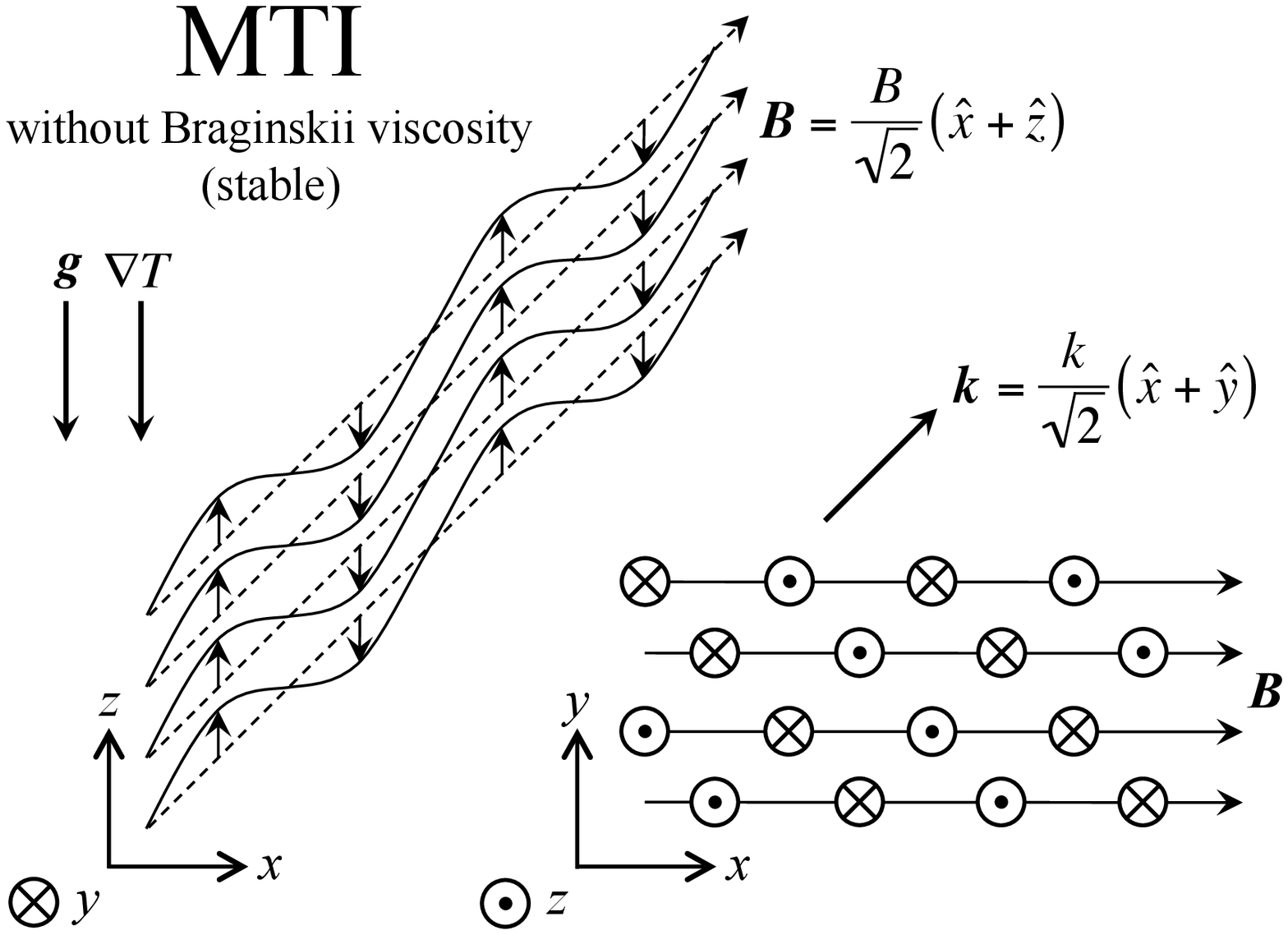}
\newline\newline\newline
\includegraphics[angle=90,width=3.2in]{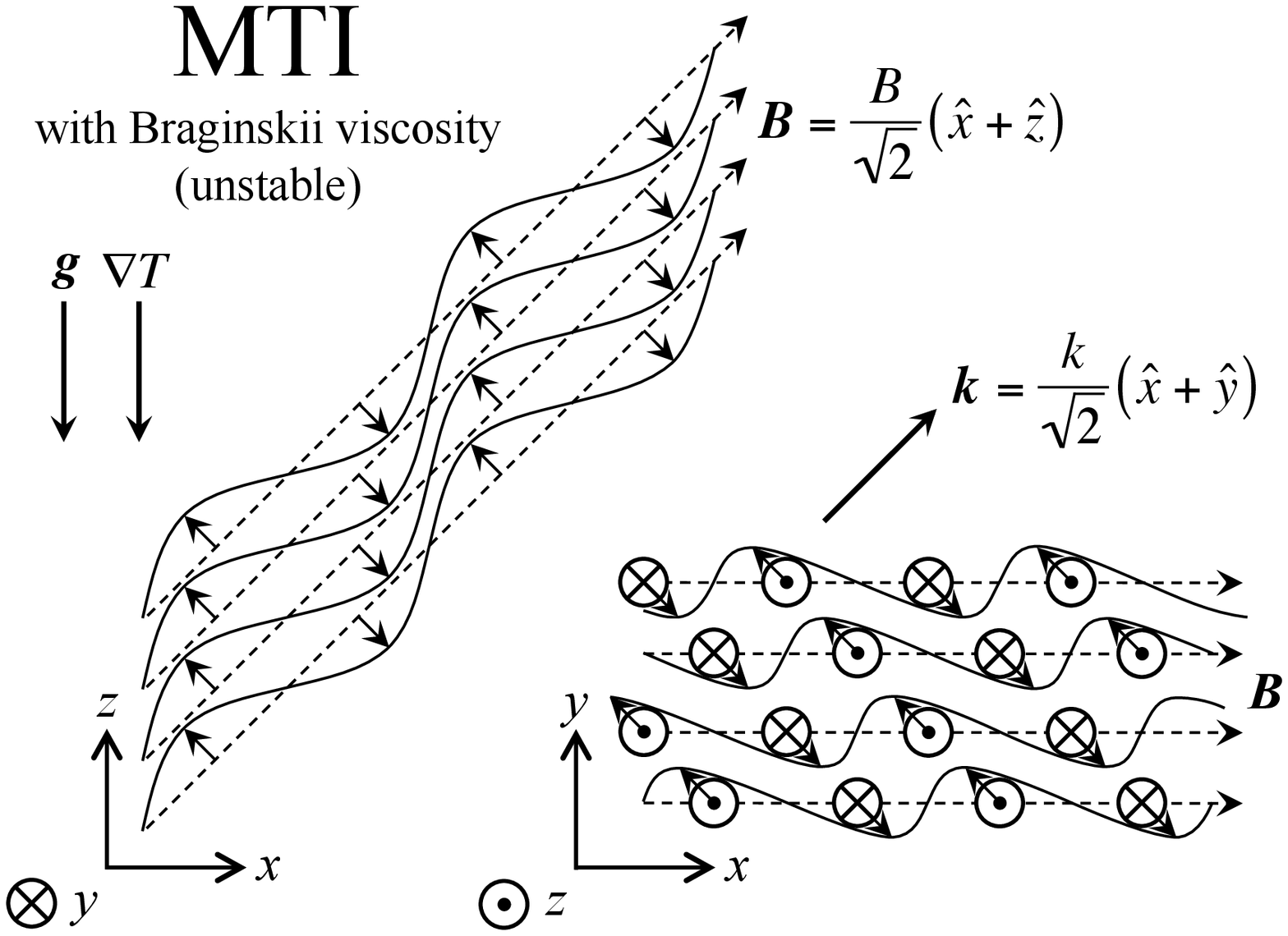}
\newline
\caption{({\it Top panel}) A stable MTI mode, with Braginskii viscosity ignored, viewed from the side (solid lines) and from above (dots and crosses). The background magnetic field $\bb{B}$ (dashed lines) makes an angle of 45 degrees with respect to both gravity and the temperature gradient. The field is perturbed with a wavevector that has equal components in the plane perpendicular to gravity, as indicated by the vector $\bb{k}$. The corresponding eigenvector has $\delta B_x = \delta B_y = 0$. This mode is MTI stable because the destabilising transfer of entropy from one fluid to another is exactly offset by the stabilising exchange of entropy by the convergence/divergence of the background heat flux. ({\it Bottom panel}) The same mode becomes unstable when Braginskii viscosity is self-consistently included. Rapid parallel viscous damping effectively orients the perturbed magnetic field nearly perpendicular to the background field, thereby precluding any stabilisation by the background heat flux. The perturbed magnetic field has components in all three directions in order to satisfy the divergence-free constraint (namely, $\delta B_z \approx \delta B_y = -\delta B_x$).}
\label{fig:mti}
\end{figure}

Matters are slightly more complicated with the MTI, which owes its existence to the alignment of isothermal magnetic field lines with the background temperature gradient. It is this alignment that allows a downwardly (upwardly) displaced fluid element always to be cooler (warmer) than the surroundings it is passing through, even as its temperature rises (falls). As the separation between magnetically-connected fluid elements grows, they take the magnetic field lines with them, aligning these heat conduits ever more parallel to the background temperature gradient and reinforcing fluid displacements. There is a weak preference for perturbations whose wavevectors are aligned with the background magnetic field; otherwise the consequent convergence (divergence) of any background heat flux would heat (cool) downwardly (upwardly) displaced fluid elements (the exact opposite of what happens with the HBI), thereby undermining the destabilising upward entropy transfer between magnetically-connected fluid elements. 

One effect of pressure anisotropy is to reinforce this preference by suppressing motions along field lines. Another potentially more important effect is to destabilise many wave modes that were previously thought to be stable to the MTI. One example is shown in Fig. \ref{fig:mti}. The panel on the top exhibits a mode that is stable to the standard MTI: the destabilising effect of generating a heat flux along field lines is exactly offset by the stabilising effect of converging/diverging background heat flux. When Braginskii viscosity is included and $\visc \gg \dyn$, the same exact mode is unstable with a growth rate $\sim$$\dyn$ (bottom panel). Any motion along the background magnetic field is rapidly damped and so the background heat flux cannot interfere with the MTI. In regions where the background heat flux converges in the $x$-$z$ plane it diverges in the $x$-$y$ plane, giving no net heat extraction. In effect, rapid Braginskii damping effectively endows slow-mode perturbations (which are subject to buoyancy forces) with Alfv\'{e}nic characteristics (i.e. perturbed magnetic fields and velocities that are predominantly oriented perpendicular to the background magnetic field).

In either case, the fundamental point is that Braginskii viscosity always suppresses perturbations that acquire free energy from the temperature gradient via a background heat flux (see eq. \ref{eqn:DT}). This is generally bad for the HBI and good for the MTI. In the next Section we formalise these qualitative expectations.

\section{Formulation of the problem}\label{sec:formulation}

\subsection{Basic Equations}\label{sec:equations}

The fundamental equations of motion are the continuity equation
\begin{equation}\label{eqn:continuity}
\D{t}{\rho} = -\rho \grad \bcdot \bb{v} ,
\end{equation}
the momentum equation
\begin{equation}\label{eqn:force}
\D{t}{\bb{v}} = -\frac{1}{\rho} \grad \bcdot \left( \msb{P} + \msb{I} \,\frac{B^2}{8\pi} - \frac{\bb{B}\bb{B}}{4\pi} \right) + \bb{g} ,
\end{equation}
and the magnetic induction equation
\begin{equation}\label{eqn:induction}
\D{t}{\bb{B}} = \bb{B} \bcdot \grad \bb{v} - \bb{B} \grad \bcdot \bb{v} ,
\end{equation}
where $\rho$ is the mass density, $\bb{v}$ the velocity, $\bb{B}$ the magnetic field, and $\bb{g}$ the gravitational acceleration; ${\rm d}/{\rm d}t \equiv \partial / \partial t + \bb{v} \bcdot \grad$ is the convective (Lagrangian) derivative. 

In the momentum equation (\ref{eqn:force}), the sum of the ion and electron pressures
\begin{equation}\label{eqn:ptensor}
\msb{P} = p_\perp \, \msb{I} - \left( p_\perp - p_{||} \right) \eb \eb
\end{equation}
is a diagonal tensor whose components perpendicular ($p_\perp$) and parallel ($p_{||}$) to the background magnetic field direction $\eb \equiv \bb{B} / B$ are in general different. Differences between the perpendicular and parallel pressure in a magnetized plasma arise from the conservation of the first and second adiabatic invariants for each particle on timescales much greater than the inverse of the cyclotron frequency. When the magnetic field strength and/or the density change, $p_\perp$ and $p_{||}$ change in different ways \citep{cgl56}. For example, conservation of the first adiabatic invariant $\mu = m v^2_\perp / 2 B$ implies that an increase in magnetic field strength must be accompanied by a corresponding increase in the perpendicular pressure, $p_\perp / B \sim {\rm const}$. 

When the collision frequency is larger than the rates of change of all fields (i.e. $\nu \gg {\rm d} / {\rm d} t$) -- a condition easily satisfied for buoyancy instabilities in the ICM -- it is straightforward to obtain an equation for the pressure anisotropy (e.g. see \citealt{scrr10} for a simple derivation):
\begin{equation}\label{eqn:anisotropy}
p_\perp - p_{||} = \frac{3 p_{\rm i}}{\nu_{\rm i}} \D{t}{} \ln \frac{B}{\rho^{2/3}} = \frac{3 p_{\rm i}}{\nu_{\rm i}} \left( \eb \eb \, \bb{:} \grad \bb{v} - \frac{1}{3} \grad \bcdot \bb{v} \right) ,
\end{equation}
where $p = (2/3) p_\perp + (1/3) p_{||}$ is the total plasma pressure.\footnote{The numerical prefactor in equation (\ref{eqn:anisotropy}) depends on the exact form of the collision operator used. A more precise numerical prefactor is $3075/1068 \simeq 2.88$ \citep[e.g.][]{cs04}; the ion collision frequency is $\nu_{\rm i} = 4 \sqrt{\pi} n_{\rm i} e^4 \Lambda_{\rm i} / 3 m^{1/2}_{\rm i} ( k_{\rm B} T_{\rm i} )^{3/2}$, where $\Lambda_{\rm i}$ is the ion Coulomb logarithm.} To obtain the final equality, we have used equations (\ref{eqn:continuity}) and (\ref{eqn:induction}) to express the rates of change of the magnetic field strength and density in terms of velocity gradients. This is referred to as the \citet{braginskii65} anisotropic viscosity. The ion contribution to the Braginskii viscosity dominates over that of the electrons by a factor $\sim$$( m_{\rm i} / m_{\rm e} )^{1/2}$. 

We will also require an internal energy equation,
\begin{equation}\label{eqn:entropy}
\frac{3}{2} \, p \, \D{t}{} \ln \frac{p}{\rho^{5/3}} = \left( p_\perp - p_{||} \right) \D{t}{} \ln \frac{B}{\rho^{2/3}} - \grad \bcdot \left( \eb\, Q \right) ,
\end{equation}
where
\begin{equation}\label{eqn:heatflux}
Q = -\chi_{\rm e} \, \eb \bcdot \grad T
\end{equation}
is the collisional heat flux, $T$ is the temperature, and $\chi_{\rm e}$ is the thermal conductivity of the electrons
\begin{equation}\label{eqn:chie}
\chi_{\rm e} \simeq 6 \times 10^{-7} \, T^{5/2}~{\rm ergs~cm}^{-1}~{\rm K}^{-1} ;
\end{equation}
the thermal conductivity of the ions is a factor $\sim$$(m_{\rm e}/m_{\rm i})^{1/2}$ smaller \citep{spitzer62}.\footnote{This assumes equal ion and electron temperatures, an assumption that may not hold in the outermost regions of galaxy clusters.} Equation (\ref{eqn:heatflux}) expresses the fact that, in the presence of a magnetic field, heat is restricted to flow along magnetic lines of force when the particle gyroradius is much smaller than the mean free path between collisions \citep[e.g.][]{braginskii65}.

\subsection{Background equilibrium and perturbations}\label{sec:equilibrium}

For simplicity, we consider a plasma stratified in both density and temperature in the presence of a uniform gravitational field in the vertical direction, $\bb{g} = -g \ez$. The plasma is not self-gravitating, so that $\bb{g}$ is a specified function of position. Without loss of generality, the magnetic field is oriented along $\eb = b_x \ex + b_z \ez$. We take the background pressure to be isotropic and $T_{\rm i} = T_{\rm e} = T$, so that $p_{\rm i} = p_{\rm e} = p / 2$. We further assume that the ratio of the ion thermal and magnetic pressures is large:
\begin{equation}
\beta \equiv \frac{8 \pi p_{\rm i}}{B^2} = \frac{\vthsq}{\vasq} \gg 1 ,
\end{equation}
where $v_{\rm th} \equiv ( p / \rho )^{1/2} = ( 2 k_{\rm B} T / m_{\rm i} )^{1/2}$ and $v_{\rm A} \equiv B / (4 \pi \rho)^{1/2}$ are the thermal and Alfv\'{e}n speeds of the ions, respectively. Observations of synchrotron radiation, inverse Compton emission, and Faraday rotation suggest a plasma $\beta$ parameter that ranges from $\sim$$10^2$ at the centres of cool-core clusters to $\sim$$10^4$ in the outermost regions of the ICM (for a review, see \citealt{ct02}). Force balance then implies
\begin{equation}
\D{z}{\ln p} = - \frac{g}{\vthsq} ,
\end{equation}
so that the inverse of the ion sound-crossing time across a thermal-pressure scale-height is equal to the dynamical frequency: 
\begin{equation}
\omega_{\rm dyn} \equiv \left( \frac{g}{H} \right)^{1/2} = \frac{v_{\rm th}}{H} .
\end{equation}

In general $\eb \bcdot \grad T \ne 0$, and so there may be a heat flux in the background state. In order to ensure our background state is in equilibrium, we must formally assume $\eb \bcdot \grad Q = 0$.\footnote{Another approach \citep[see][]{br10} is to construct an equilibrium state in which conductive heating is balanced by radiative cooling.} However, as long as the timescale for the evolution of the background (global) state is longer than the local dynamical time, our results do not depend critically on the system actually being in global steady state.

We allow perturbations (denoted by a $\delta$) about the background state and order their amplitudes as follows:
\begin{equation}\label{eqn:ampordering}
\frac{\delta v}{v_{\rm th}} \sim \frac{\delta \rho}{\rho} \sim \frac{\delta T}{T} \sim \frac{1}{\mc{M}} \frac{\delta p}{p} \sim \frac{1}{\beta^{1/2}} \frac{\delta B}{B} \sim \frac{1}{\beta^{1/2}} \sim \mc{M} ,
\end{equation}
where $\mc{M} \ll 1$ is the Mach number. This amounts to the Boussinesq approximation (i.e. relative changes in the pressure are much smaller than relative changes in the temperature or density). Sound waves are then eliminated from the analysis and so the flow behaves as though it were incompressible, a good approximation in the ICM where typical velocities are much smaller than the sound speed.

The perturbations are taken to have space-time dependence $\exp( \sigma t + {\rm i} \bb{k} \bcdot \bb{r} )$, where the growth rate $\sigma$ may be complex and the wavevector $\bb{k} = k_x \ex + k_y \ey + k_z \ez$. We order the timescales and the spatial scales as follows: 
\begin{equation}
\sigma \sim \dyn \sim \cond \sim \visc \sim k v_{\rm A} \sim \mc{M} \, k v_{\rm th} \sim \mc{M}^2 \nu_{\rm i} ,
\end{equation}
\begin{equation}
k \sim \frac{1}{\mc{M} H} \sim \frac{\mc{M}}{\mfp} .
\end{equation}
The latter ordering means that the relevant wavelengths are intermediate between micro- and macroscopic, viz. $k ( \mfp H )^{1/2} \sim 1$. Note that we are formally treating $( m_{\rm e} / m_{\rm i} )^{1/2}$ as a parameter of order unity here -- a subsidiary expansion with respect to it will be done later. This completes the formulation of the problem.

\subsection{Linearised equations}\label{sec:lineareqns}

With account taken of the ordering introduced in Section \ref{sec:equilibrium}, the linearised versions of equations (\ref{eqn:continuity})--(\ref{eqn:induction}) and (\ref{eqn:anisotropy})--(\ref{eqn:heatflux}) are then
\begin{equation}\label{eqn:lin:continuity}
\bb{k} \bcdot \delta \bb{v} = 0 ,
\end{equation}
\begin{eqnarray}\label{eqn:lin:force}
\lefteqn
{
\sigma \delta \bb{v} = -{\rm i} \bb{k} \, \vthsq \left( \frac{\delta p_\perp}{p} + \frac{1}{\beta} \frac{\delta B_{||}}{B} \right) + {\rm i} k_{||} \, \vasq \frac{\delta\bb{B}}{B} - g \ez \,\frac{\delta\rho}{\rho} 
} 
\nonumber\\*&& \mbox{} 
- \eb \, \frac{3}{2} \frac{k^2_{||} \vthsq}{\nu_{\rm i}} \, \delta v_{||} ,
\end{eqnarray}
\begin{equation}\label{eqn:lin:induction}
\sigma \delta \bb{B} = {\rm i} k_{||} B \delta \bb{v} ,
\end{equation}
\begin{equation}\label{eqn:lin:entropy}
\sigma \frac{\delta \rho}{\rho} - \delta v_z \, \frac{3}{5} \D{z}{} \ln \frac{p}{\rho^{5/3}} = \frac{2{\rm i}} {5p} \, \bb{k} \bcdot \left( \eb \, \delta Q - \delta \eb \, \chi_{\rm e} b_z \D{z}{T} \right) ,
\end{equation}
\begin{equation}\label{eqn:lin:heatflux}
\delta Q = -\chi_{\rm e} \, \delta b_z \D{z}{T} - \chi_{\rm e} \, {\rm i} k_{||} \delta T ,
\end{equation}
\begin{equation}\label{eqn:lin:pressurebalance}
\frac{\delta\rho}{\rho} = -\frac{\delta T}{T} ,
\end{equation}
where the subscript $||$ denotes the vector component parallel to the background magnetic field (e.g. $k_{||} = \eb \bcdot \bb{k}$) and $\delta \eb = \delta \bb{B}_\perp / B$ is the perturbation of the unit vector $\eb$. Equation (\ref{eqn:lin:pressurebalance}) expresses pressure balance for the perturbations, a consequence of our low-Mach-number ordering (eq. \ref{eqn:ampordering}). The total perpendicular pressure perturbation $\delta p_\perp$ in equation (\ref{eqn:lin:force}) is found by enforcing incompressibility (eq. \ref{eqn:lin:continuity}). 

The linearised entropy equation (\ref{eqn:lin:entropy}) deserves special attention. The first term on the right-hand side is responsible for the MTI. If the temperature {\em increases} in the direction of gravity, any alignment between the perturbed magnetic field direction and the temperature gradient ($\delta b_z \ne 0$) is unstable as long as conduction is rapid enough to ensure approximately isothermal field lines. When the temperature {\em decreases} in the direction of gravity, this term is stabilising. The second term on the right-hand side is responsible for the HBI. If the temperature {\em decreases} in the direction of gravity, convergence/divergence of heat-flux--channeling magnetic field lines ($\bb{k} \bcdot \delta \eb \ne 0$) leads to buoyantly-unstable density perturbations. When the temperature {\em increases} in the direction of gravity, this term is stabilising .

Using equations (\ref{eqn:lin:pressurebalance}) and (\ref{eqn:lin:heatflux}), the linearised internal energy equation (\ref{eqn:lin:entropy}) becomes
\begin{eqnarray}\label{eqn:lin:entropy2}
\lefteqn
{
\left( \sigma + \cond \right) \frac{\delta \rho}{\rho} = \delta v_z \, \frac{3}{5} \D{z}{} \ln \frac{p}{\rho^{5/3}}
}
\nonumber\\*&& \mbox{} 
- {\rm i} \, \cond \, \frac{\delta B_z - 2 b_z \delta B_{||}}{B} \frac{1}{k_{||}} \D{z}{\ln T}
\end{eqnarray}
to leading order in $\mc{M}$. Here we have introduced the characteristic conduction frequency,
\begin{eqnarray}\label{eqn:wcond}
\cond & \equiv & \frac{2}{5} k^2_{||} \frac{\chi_{\rm e} T}{p} = \frac{3.2}{10} \frac{\Lambda_{\rm i}}{\Lambda_{\rm e}} \left( \frac{m_{\rm i}}{2 m_{\rm e}} \right)^{1/2}  k^2_{||} \mfp H \, \dyn
\nonumber\\*
\mbox{} & \approx & 10 \, k^2_{||} \mfp H \, \dyn ,
\end{eqnarray}
where $\Lambda_{\rm i}$ ($\Lambda_{\rm e}$) is the Coulomb logarithm of the ions (electrons).

Equations (\ref{eqn:lin:continuity})--(\ref{eqn:lin:entropy2}) differ from those in \citet{quataert08} only by the final term in the momentum equation (\ref{eqn:lin:force}), which is due to the perturbed pressure anisotropy (Braginskii viscosity). This term introduces a characteristic frequency associated with viscous damping:
\begin{equation}\label{eqn:wvisc}
\visc \equiv \frac{3}{2} \frac{k^2_{||} \vthsq}{\nu_{\rm i}}  = \frac{3}{2} \, k^2_{||} \mfp H \, \omega_{\rm dyn} ,
\end{equation}
which is a factor $\approx$$6$ smaller than $\cond$.

\subsection{Dispersion Relation}\label{sec:disprel}

The dispersion relation that results after combining equations (\ref{eqn:lin:continuity})--(\ref{eqn:lin:induction}) and (\ref{eqn:lin:entropy2}) may be written in the following form:
\begin{eqnarray}\label{eqn:disprel1}
\lefteqn
{
-\visc \frac{k^2_\perp}{k^2}
}
\nonumber\\*&&\mbox{}
= { {\displaystyle \widetilde{\sigma}^2 \left[ \widetilde{\sigma}^2 \left( \sigma + \cond \right) + \sigma N^2 \frac{k^2_x+k^2_y}{k^2} + \cond \, g \D{z}{\ln T} \frac{\mc{K}}{k^2} \right]} \over {\displaystyle \sigma \left[ \widetilde{\sigma}^2 \left( \sigma + \cond \right) + \sigma N^2 \frac{b^2_x k^2_y}{k^2_\perp} + \cond \, g \D{z}{\ln T} \frac{b^2_x k^2_y}{k^2_\perp} \right]}} ,
\nonumber\\*
\end{eqnarray}
where $k^2_\perp \equiv k^2 - k^2_{||}$ is the square of the wavevector component perpendicular to the background magnetic field,\footnote{In contrast to our notation, \citet{quataert08} and \citet{br10} use $k^2_\perp$ to denote the square of the wavevector component perpendicular to {\em gravity}, not to the background magnetic field.}
\begin{equation}
\widetilde{\sigma}^2 \equiv \sigma^2 + k^2_{||} \vasq ,
\end{equation}
and
\begin{eqnarray}\label{eqn:mck}
\mc{K} &\equiv & \left(1 - 2 b^2_z\right) \left(k^2_x + k^2_y\right)  + 2 b_x b_z k_x k_z \nonumber\\*
 &=& b^2_x k^2 - k^2_\perp + b^2_x k^2_y = - b^2_z k^2 + k^2_{||} + b^2_x k^2_y .
\end{eqnarray}
We have written $\mc{K}$ in three equivalent forms, all of which will prove useful in our analysis. We have also introduced the Brunt-V\"{a}is\"{a}l\"{a} frequency given by
\begin{equation}
N^2 \equiv \frac{3}{5} g \D{z}{} \ln \frac{p}{\rho^{5/3}} > 0 .
\end{equation}
Were conduction, pressure anisotropy, and the magnetic field all to be ignored, equation (\ref{eqn:disprel1}) would reduce to the usual dispersion relation for internal gravity waves, $\sigma^2 = - N^2 ( k^2_x + k^2_y ) / k^2$. 

It will also be beneficial to have the equations for the perturbations at hand, written in the limit of fast conduction ($\cond \gg \dyn \sim \sigma$):
\begin{eqnarray}\label{eqn:eigenrho}
\lefteqn
{
\frac{\delta \rho}{\rho} = - \frac{\delta T}{T} \simeq \xi_z \D{z}{\ln T}
}
\nonumber\\*&&
\mbox{}  \times \left[ { { \displaystyle \widetilde{\sigma}^2 ( k_z - 2 b_z k_{||} ) - \sigma \visc \, b_x ( \eb \btimes \bb{k} )_y } \over { \displaystyle \widetilde{\sigma}^2 k_z - \sigma \visc \, b_x ( \eb \btimes \bb{k} )_y + g \D{z}{\ln T} 2 b_x b_z k_x } } \right] ,
\nonumber\\ &&
\end{eqnarray}
\begin{eqnarray}\label{eqn:eigenbx}
\lefteqn
{
\frac{\delta B_x}{B} = {\rm i} k_{||} \xi_x \simeq {\rm i} k_{||} \xi_z
}
\nonumber\\*&&
\mbox{} \times \left[ { { \displaystyle \widetilde{\sigma}^2 k_x + \sigma \visc \, b_z ( \eb \btimes \bb{k} )_y + g \D{z}{\ln T} ( 1 - 2 b^2_z ) k_x } \over { \displaystyle \widetilde{\sigma}^2 k_z - \sigma \visc \, b_x ( \eb \btimes \bb{k} )_y + g \D{z}{\ln T} 2 b_x b_z k_x } } \right] ,
\nonumber\\*
\end{eqnarray}
\begin{eqnarray}\label{eqn:eigenby}
\lefteqn
{
\frac{\delta B_y}{B} = {\rm i} k_{||} \xi_y \simeq {\rm i} k_{||} \xi_z \frac{1}{k_y}
}
\nonumber\\*&&
\mbox{} \times  \left[ { { \displaystyle \widetilde{\sigma}^2 k^2_y + \sigma \visc \, k^2_y + g \D{z}{\ln T} ( 1 - 2 b^2_z ) k^2_y } \over { \displaystyle \widetilde{\sigma}^2 k_z - \sigma \visc \, b_x ( \eb \btimes \bb{k} )_y + g \D{z}{\ln T} 2 b_x b_z k_x } } \right.
\nonumber\\*&&
\left. \mbox{} -  { { \displaystyle \widetilde{\sigma}^2 k^2 + \sigma \visc \, k^2_\perp + g \D{z}{\ln T} \mc{K} } \over { \displaystyle \widetilde{\sigma}^2 k_z - \sigma \visc \, b_x ( \eb \btimes \bb{k} )_y + g \D{z}{\ln T} 2 b_x b_z k_x } }\right] ,
\end{eqnarray}
\begin{equation}\label{eqn:eigenbz}
\frac{\delta B_z}{B} = {\rm i} k_{||} \xi_z ,
\end{equation}
\begin{eqnarray}\label{eqn:eigenb}
\lefteqn
{
\frac{\delta B_{||}}{B} =  {\rm i} k_{||} \xi_{||} \simeq {\rm i} k_{||} \xi_z 
}
\nonumber\\*&&
\mbox{} \times \left[ { {\displaystyle \widetilde{\sigma}^2 k_{||} + g \D{z}{\ln T} b_x k_x } \over {\displaystyle \widetilde{\sigma}^2 k_z - \sigma \visc \, b_x ( \eb \btimes \bb{k} )_y + g \D{z}{\ln T} 2 b_x b_z k_x } } \right] ,
\end{eqnarray}
\begin{eqnarray}\label{eqn:eigenbprp}
\lefteqn
{
\frac{\delta \bb{B}_\perp}{B} = {\rm i} k_{||} \bb{\xi}_\perp \simeq \frac{\delta B_y}{B} \, \ey + {\rm i} k_{||} \xi_z \, ( \ey \btimes \eb )
}
\nonumber\\*&&
\mbox{} \times \left[ { {\displaystyle \widetilde{\sigma}^2 ( \eb \btimes \bb{k} )_y + \sigma \visc ( \eb \btimes \bb{k} )_y - g \D{z}{\ln T} b_z k_x } \over {\displaystyle \widetilde{\sigma}^2 k_z - \sigma \visc \, b_x ( \eb \btimes \bb{k} )_y + g \D{z}{\ln T} 2 b_x b_z k_x } }\right] ,
\end{eqnarray}
where $\bb{\xi} = \sigma \delta \bb{v}$ is the Lagrangian displacement of a fluid element ($\xi_z > 0$ is upward). Equations (\ref{eqn:eigenrho}) and (\ref{eqn:eigenb}) imply that the Lagrangian change in the temperature of a fluid element is
\begin{eqnarray}\label{eqn:DT}
\lefteqn
{
\frac{\Delta T}{T} = \frac{\delta T}{T} + \xi_z \D{z}{\ln T} \simeq 2 b_z \xi_{||} \D{z}{\ln T} = -\frac{2 {\rm i} b_z}{k_{||}} \D{z}{\ln T} \frac{\delta B_{||}}{B} ,
}
\nonumber\\* 
\end{eqnarray}
emphasizing that {\em perturbations in magnetic-field strength go hand-in-hand with changes in temperature}. This will turn out to be an extremely important property for understanding the results of Section \ref{sec:results}.

\subsection{Nature of perturbations}

If we set $\visc = 0$, equation (\ref{eqn:disprel1}) returns the standard MTI-HBI dispersion relation \citep[see eq. 13 of][]{quataert08}:
\begin{eqnarray}\label{eqn:mtihbi}
\lefteqn
{ 
\widetilde{\sigma}^2 \left[ \widetilde{\sigma}^2 \left(\sigma + \cond\right) + \sigma N^2 \frac{k^2_x + k^2_y}{k^2} + \cond \, g \D{z}{\ln T} \frac{\mc{K}}{k^2} \right] = 0 .
}
\nonumber\\*
\end{eqnarray}
The $\widetilde{\sigma}^2 = 0$ branch of this dispersion relation represents Alfv\'{e}n waves that are polarised with $\delta \bb{B}$ along the $y$-axis. They are unaffected by buoyancy. The other three modes are coupled slow and entropy modes. If we further take the limit of fast conduction ($\cond \gg \dyn \sim \sigma$), the entropy mode becomes $\sigma \simeq -\cond$, while the slow modes satisfy
\begin{equation}\label{eqn:mtihbi2}
\widetilde{\sigma}^2 \simeq -g \, \D{z}{\ln T} \frac{\mc{K}}{k^2} .
\end{equation}
When the temperature gradient and $\mc{K}$ have opposite signs, one of the slow modes may become unstable to the MTI/HBI.

Braginskii viscosity modifies this picture in two ways. First consider the limit $k_y = 0$, in which the wavevector lies entirely in the plane spanned by gravity and the background magnetic field. In this case, equation (\ref{eqn:disprel1}) becomes
\begin{eqnarray}\label{eqn:brag1}
\lefteqn
{
\widetilde{\sigma}^2 \left[ \widetilde{\sigma}^2 \left( \sigma + \cond \right) + \sigma \visc \frac{k^2_\perp}{k^2} \left( \sigma + \cond \right) \right .
}
\nonumber\\*&& 
\left. \mbox{} + \sigma N^2 \frac{k^2_x + k^2_y}{k^2} + \cond \, g \D{z}{\ln T} \frac{\mc{K}}{k^2} \right] = 0 .
\end{eqnarray}
The Alfv\'{e}n-wave branch of the dispersion relation is unchanged, since Braginskii viscosity does not affect motions perpendicular to the magnetic field. In contrast, slow-mode--polarised perturbations with $k_\perp \ne 0$ are damped. This property of Braginskii viscosity is the root cause of the significant changes to the nature of the allowed unstable HBI (Section \ref{sec:standardhbi}) and MTI (Section \ref{sec:standardmti}) modes.

Next consider the general dispersion relation (\ref{eqn:disprel1}) with $k_y \ne 0$. In this case, {\em Braginskii viscosity couples the Alfv\'{e}n and slow modes}. This can be seen most clearly by taking the fast-conduction limit ($\cond \gg \dyn \sim \sigma$) of equation (\ref{eqn:disprel1}):
\begin{eqnarray}\label{eqn:disprel2}
\lefteqn
{
\widetilde{\sigma}^2 \left( \widetilde{\sigma}^2 + \sigma \visc \frac{k^2_\perp}{k^2} + g \D{z}{\ln T} \frac{\mc{K}}{k^2} \right) \simeq - \sigma \visc \,  g \D{z}{\ln T} \frac{b^2_x k^2_y}{k^2} .
}
\nonumber\\*
\end{eqnarray}
When $\visc \gg \dyn$ the slow mode is rapidly damped, leaving only $\sigma \visc \, k^2_\perp / k^2$ to highest order in the parentheses on the left-hand side of equation (\ref{eqn:disprel2}). This term cancels the similar factor on the right-hand side, ultimately leading to
\begin{equation}\label{eqn:brag2}
\widetilde{\sigma}^2 \simeq - g \D{z}{\ln T} \frac{b^2_x k^2_y}{k^2_\perp} ,
\end{equation}
which may be unstable when the temperature increases in the direction of gravity.\footnote{There are other instances of an anisotropic damping mechanism coupling the Alfv\'{e}n and slow mode branches of a dispersion relation via a free energy gradient. In weakly-ionised plasmas, the interaction between velocity shear and anisotropic magnetic resistivity (ambipolar diffusion and the Hall effect) results in such a coupling -- one which ultimately leads to shear-driven instabilities \citep{kunz08}.} We will elaborate on this result in Section \ref{sec:alfvenicmti}, where we discuss this new `Alfv\'{e}nic' version of the MTI, but for now we explain the physical content of equations (\ref{eqn:disprel2}) and (\ref{eqn:brag2}). By damping motions along field lines, Braginskii viscosity effectively reorients magnetic field perturbations to be nearly perpendicular to the background magnetic field (via flux freezing). These modes therefore display characteristics of both slow and Alfv\'{e}n modes: they have density and temperature perturbations, and therefore are subject to buoyancy forces, but their velocity and magnetic field perturbations are predominantly polarised across the mean field.

We note in passing the striking similarity between equation (\ref{eqn:disprel2}) and the dispersion relation for the axisymmetric magnetorotational instability subject to Braginskii stresses \citep{balbus04,ib05}:
\begin{eqnarray}\label{eqn:mrimvi}
\lefteqn
{
\widetilde{\sigma}^2 \left( \widetilde{\sigma}^2 + \sigma \visc \frac{k^2_\perp}{k^2} + g \D{R}{\ln \Omega^2} \frac{k^2_Z}{k^2} \right)
}
\nonumber\\*&&
\mbox{} = -\sigma \visc \, g \D{R}{\ln \Omega^2} \frac{b^2_\phi k^2_Z}{k^2} - 4 \Omega^2 \frac{k^2_Z}{k^2} \sigma^2
\end{eqnarray}
where $g = \Omega^2 R$ in a rotating disc. Aside from a $4 \Omega^2$ term due to epicyclic motions, the equivalence is revealed by relabelling the disc coordinate system $( R , \phi , Z )\leftrightarrow ( z , x , y )$ and swapping one free energy source (temperature gradient) for another (angular velocity gradient).\footnote{We refer the reader to \citet{balbus00,balbus01} for a cogent discussion of the analogy between angular momentum and entropy that underlies these mathematical similarities.} Braginskii viscosity couples the Alfv\'{e}n- and slow-mode branches of the dispersion relation via the angular velocity gradient in very much the same way that it coupled these branches via the temperature gradient in equation (\ref{eqn:disprel2}). Furthermore, when the angular velocity decreases outwards and $\visc \gg \Omega \sim \sigma$, the first term on the right-hand side of equation (\ref{eqn:mrimvi}) due to Braginskii viscosity overwhelms the second term due to epicyclic coupling and drives the magnetoviscous instability (MVI) by endowing slow-mode perturbations with Alfv\'{e}n-mode characteristics (as in eq. \ref{eqn:disprel2}). We therefore identify the behaviour revealed by equation (\ref{eqn:brag2}) as the temperature-gradient analog of the MVI.

\section{Results}\label{sec:results}

%
%
\begin{figure}
\centering
\includegraphics[width=2.7in]{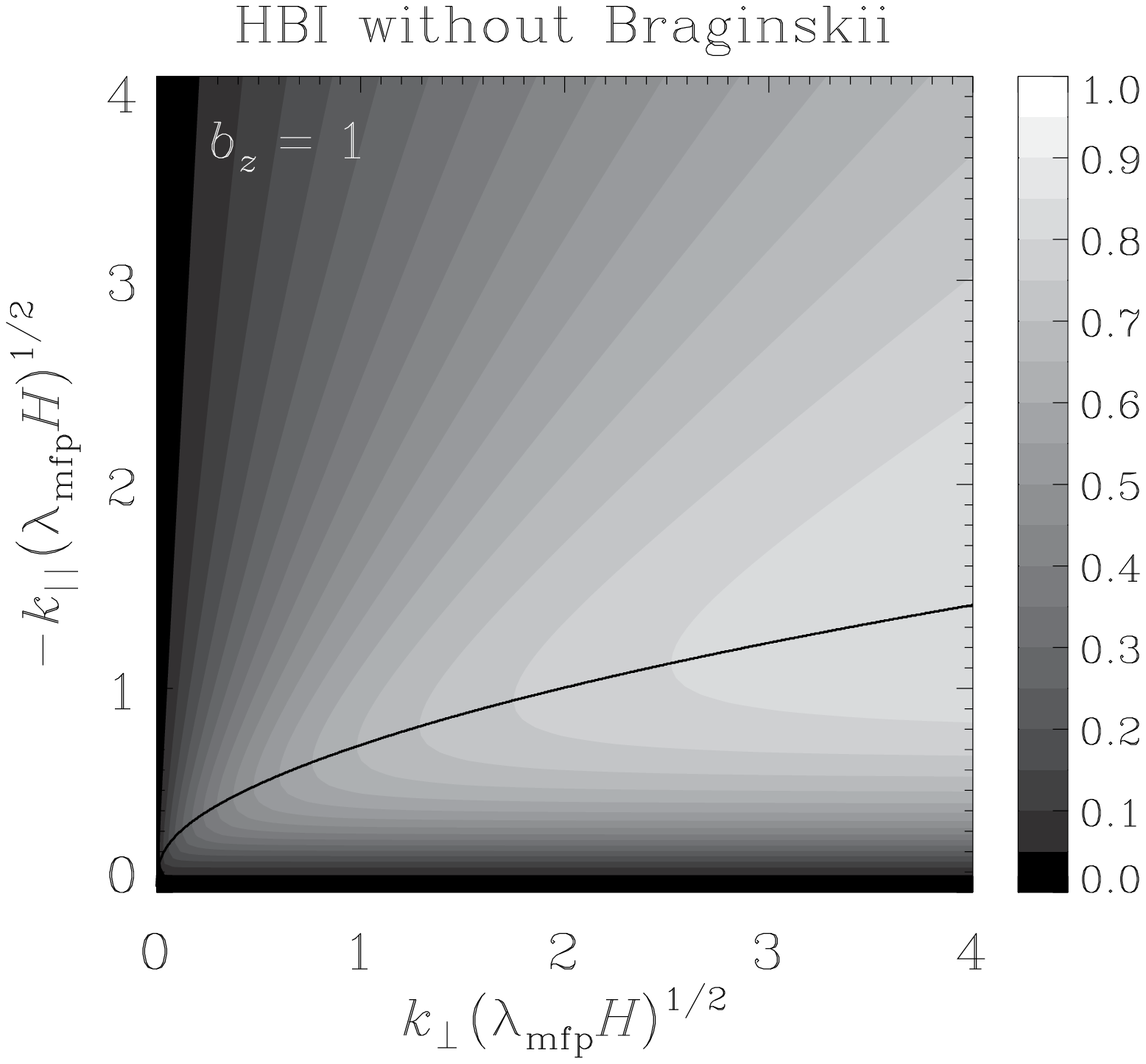}
\newline
\newline
\includegraphics[width=2.7in]{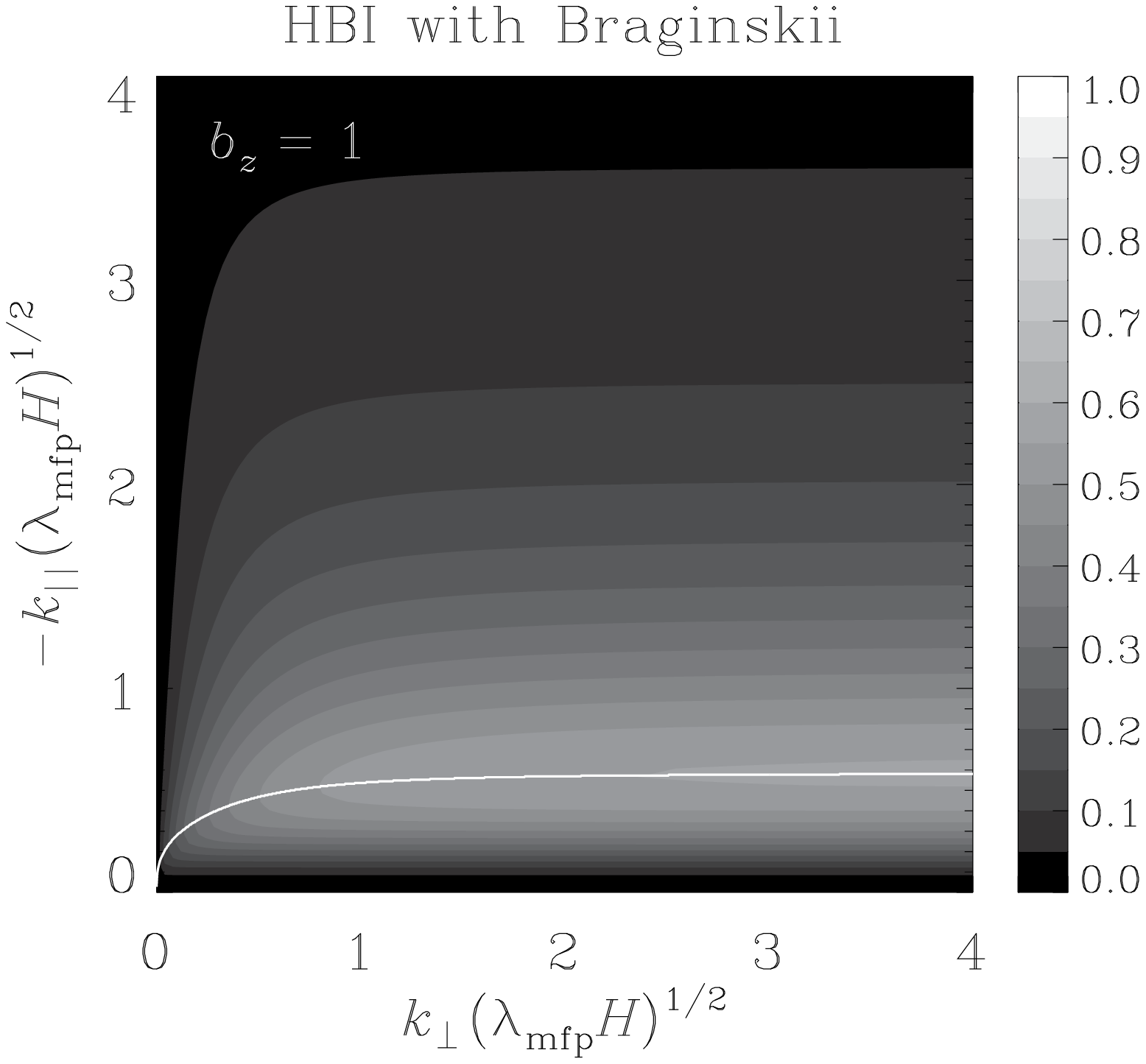}
\newline
\caption{HBI growth rate (normalised to the maximum growth rate $\sqrt{ g \, {\rm d} \ln T / {\rm d} z}$) without ({\it top}) and with ({\it bottom}) Braginskii viscosity for a stratified thermal layer with ${\rm d} \ln T / {\rm d} \ln p = -1$ threaded by a vertical magnetic field ($b_z = 1$). Magnetic tension is neglected; its effect is discussed in Section \ref{sec:hbitension}. Each contour represents an increase in the growth rate by 5 per cent. The solid lines (given asymptotically by eq. \ref{eqn:linefit} with Braginskii viscosity and eq. \ref{eqn:approxhbi} without Braginskii viscosity) trace the maximum growth rate for a given total wavenumber $k$; the maximum growth rate is given by eq. \ref{eqn:approxgmax} with Braginskii viscosity and asymptotically at $k_{||} (\mfp H)^{1/2} \gg 1$ by eq. \ref{eqn:hbimax} without Braginskii viscosity. Braginskii viscosity dramatically reduces growth rates everywhere except for a narrow band of wavenumbers around $k_{||}$ given by eq. \ref{eqn:linefit}. Galaxy clusters with cool cores typically have $H / \mfp \sim 10^2$--$10^3$ at radii for which the temperature profile increases outwards, and so the maximum wavenumbers for each axis span the range $k H \sim 40$--$127$.}
\label{fig:hbi_kxkz}
\end{figure}

\subsection{${\rm d}T/{\rm d}z > 0$: Heat-flux--driven Buoyancy Instability}\label{sec:hbi}

We first investigate the effects of Braginskii viscosity on the stability of a stratified atmosphere in which the temperature decreases in the direction of gravity, i.e. ${\rm d}T / {\rm d} z > 0$. Such an atmosphere was shown by \citet{quataert08} to be susceptible to the HBI if $\mc{K} < 0$.

\subsubsection{Case of $k_y = 0$: Standard HBI with and without Braginskii viscosity}\label{sec:standardhbi}

If Braginskii viscosity is ignored, it is straightforward to show from equation (\ref{eqn:mtihbi}) that the maximum HBI growth rate
\begin{equation}\label{eqn:hbimax}
\sigma^2_{\rm HBI,max} \simeq g \D{z}{\ln T} b^2_z
\end{equation}
occurs for wavevectors satisfying
\begin{equation}\label{eqn:hbikpkprp}
\frac{k^2_{||}}{k^2_\perp} \simeq b^2_z \, \frac{\sigma_{\rm HBI,max}}{\cond} \left( 1 + \frac{1}{5} \left| \D{\ln T}{\ln p} \right| \right) \ll 1
\end{equation}
to leading order in $\dyn / \cond$, where we have assumed $k_{||} H \ll b_z \beta^{1/2}$ -- i.e. magnetic tension is negligible on the scales of interest. Equation (\ref{eqn:hbikpkprp}) reveals that the HBI has a strong preference for perpendicular wavenumbers. More precisely, using the definition of $\cond$ (eq. \ref{eqn:wcond}) in equation (\ref{eqn:hbikpkprp}), we find that the maximum growth rate occurs along a path through $k$-space on which
\begin{eqnarray}\label{eqn:approxhbi}
\lefteqn
{
k_{||} ( \mfp H )^{1/2} \approx \pm k^{1/2}_\perp ( \mfp H )^{1/4} 
}
\nonumber\\*&&
\mbox{} \times  0.6 b^{3/4}_z \left( 1 + \frac{1}{5} \left| \D{\ln T}{\ln p} \right| \right)^{1/4} \left| \D{\ln p}{\ln T} \right|^{1/8} .
\end{eqnarray}
This behaviour is exhibited in the top panel of Fig. \ref{fig:hbi_kxkz}, which shows HBI growth rates in the $( k_{||} , k_\perp )$ plane for $b_z = 1$ (without Braginskii viscosity). The solid line in the plot traces the maximum growth rate through wavenumber space; it quickly asymptotes to equation (\ref{eqn:approxhbi}).

The HBI's preference for perpendicular wavenumbers is also reflected in the corresponding eigenvectors. Using equations (\ref{eqn:eigenrho}) and (\ref{eqn:mtihbi2}) we find that the density perturbation associated with the HBI,
\begin{equation}
\frac{\delta \rho}{\rho} \simeq -\xi_z \D{z}{\ln T} \frac{k^2}{k^2_x} \left( b^2_z - \frac{k^2_{||}}{k^2} \right) ,
\end{equation}
is greatest when $k^2_{||} \ll k^2$ so that, e.g., upwardly-displaced fluid elements have the largest possible decrease in their density. Moreover, perpendicular wavenumbers are necessary to generate linear perturbations in magnetic field strength,
\begin{equation}
\frac{\delta B_{||}}{B} \simeq {\rm i} k_{||} \xi_z \frac{k_\perp}{k_x} ,
\end{equation}
which lead to local convergence/divergence of the background heat flux and consequent heating/cooling of the plasma (eq. \ref{eqn:DT}).

The problem is that it is precisely such perturbations that are damped by Braginskii viscosity (see the bottom panel of Fig. \ref{fig:hbi_kxkz}). By equation (\ref{eqn:DT}), upward displacements along magnetic field lines go hand-in-hand with local heating ($\Delta T > 0$) and a local increase in the magnetic field strength ($\delta B_{||} > 0$).\footnote{While Alfv\'{e}nically-polarised modes suffer no viscous damping, they are HBI stable because $\delta B_{||} = 0$.} This causes a negative viscous stress that damps motions along field lines, thereby rarefying the magnetic field and reducing the strength of the perturbed heat flux. This can be seen quantitatively by explicitly writing down the buoyancy and viscous forces in the $z$-component of the momentum equation (\ref{eqn:lin:force}) for the simple case $b_z = 1$:
\begin{equation}
\DD{t}{\xi_z} = \dots + g \D{z}{\ln T} \xi_z - \visc \D{t}{\xi_z} .
\end{equation}
For wavenumbers satisfying the ordering $\visc \gg \dyn$, or $k^2_{||} \mfp H \gg 1$, it is straightforward to show from equation (\ref{eqn:disprel2}) that the growth rate is
\begin{equation}
\sigma \simeq \frac{g}{\visc} \D{z}{\ln T} \sim \frac{\dyn^2}{\visc}
\end{equation}
to leading order in $\dyn / \visc$. In other words, the buoyancy and viscous forces become nearly equal and opposite as the plasma becomes more and more collisionless. The growth rate decreases accordingly.

%
%
\begin{figure*}
\centering
\includegraphics[height=2.2in]{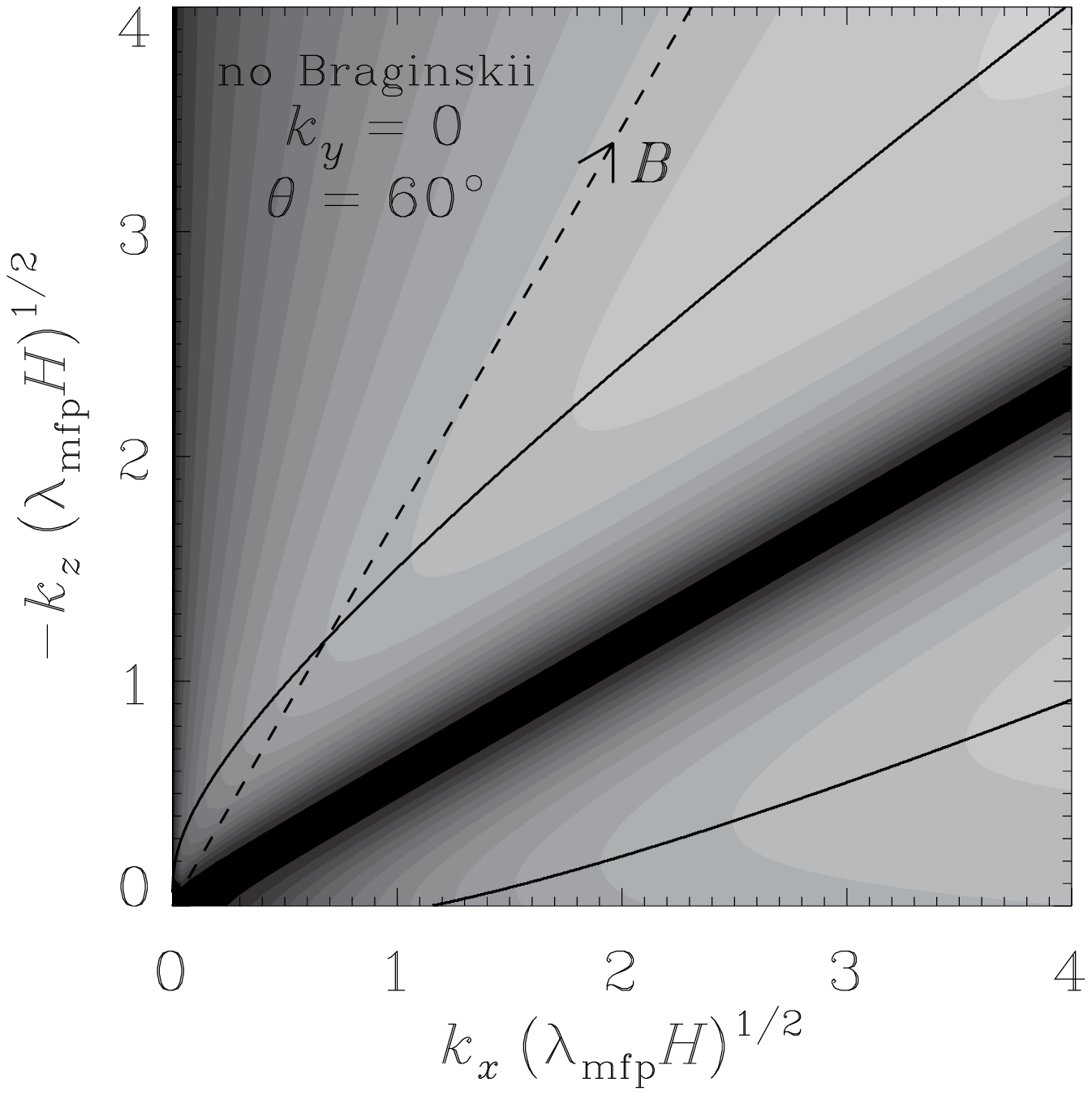}\quad
\includegraphics[height=2.2in]{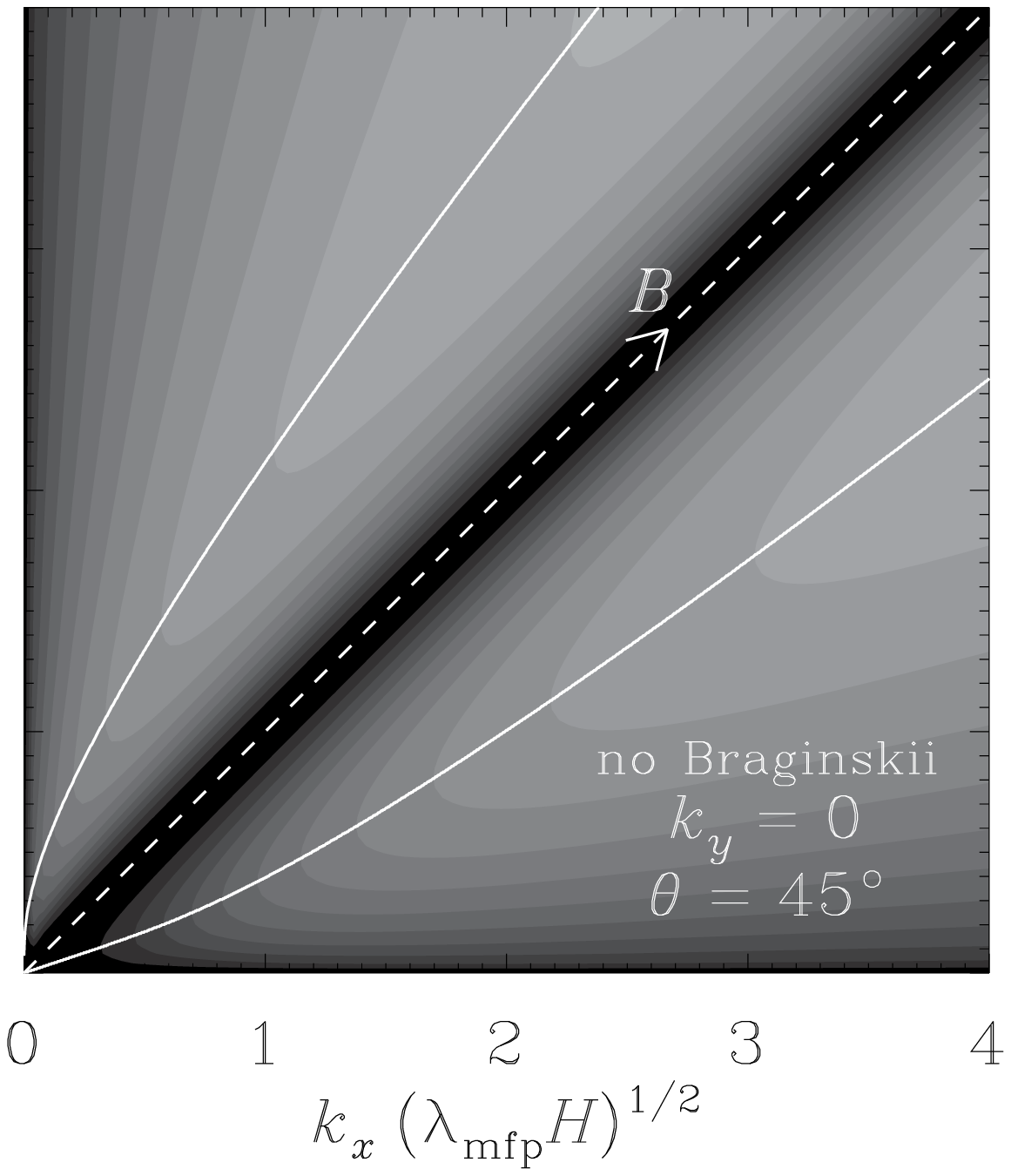}\quad
\includegraphics[height=2.2in]{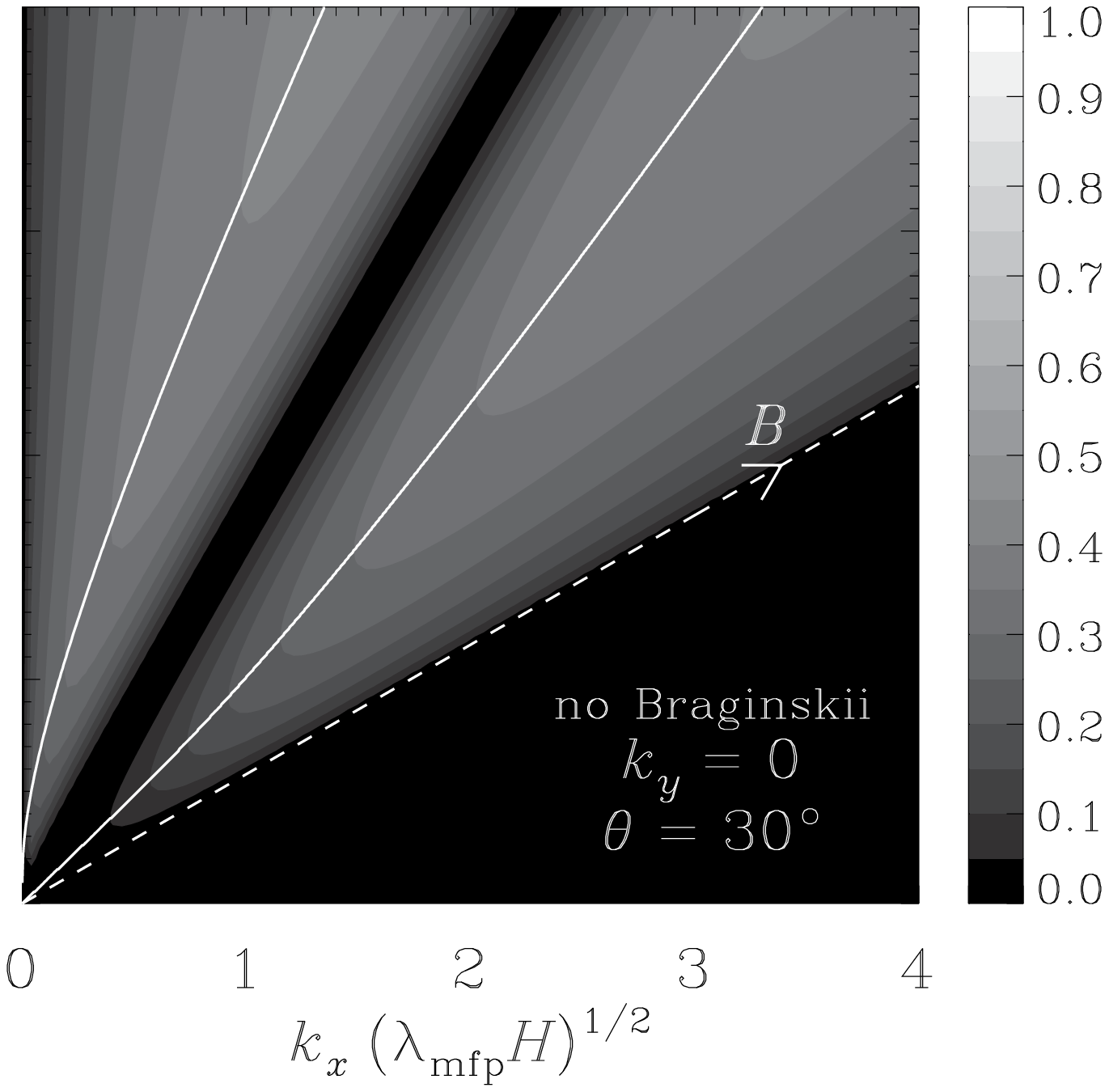}
\newline
\newline
\includegraphics[height=2.2in]{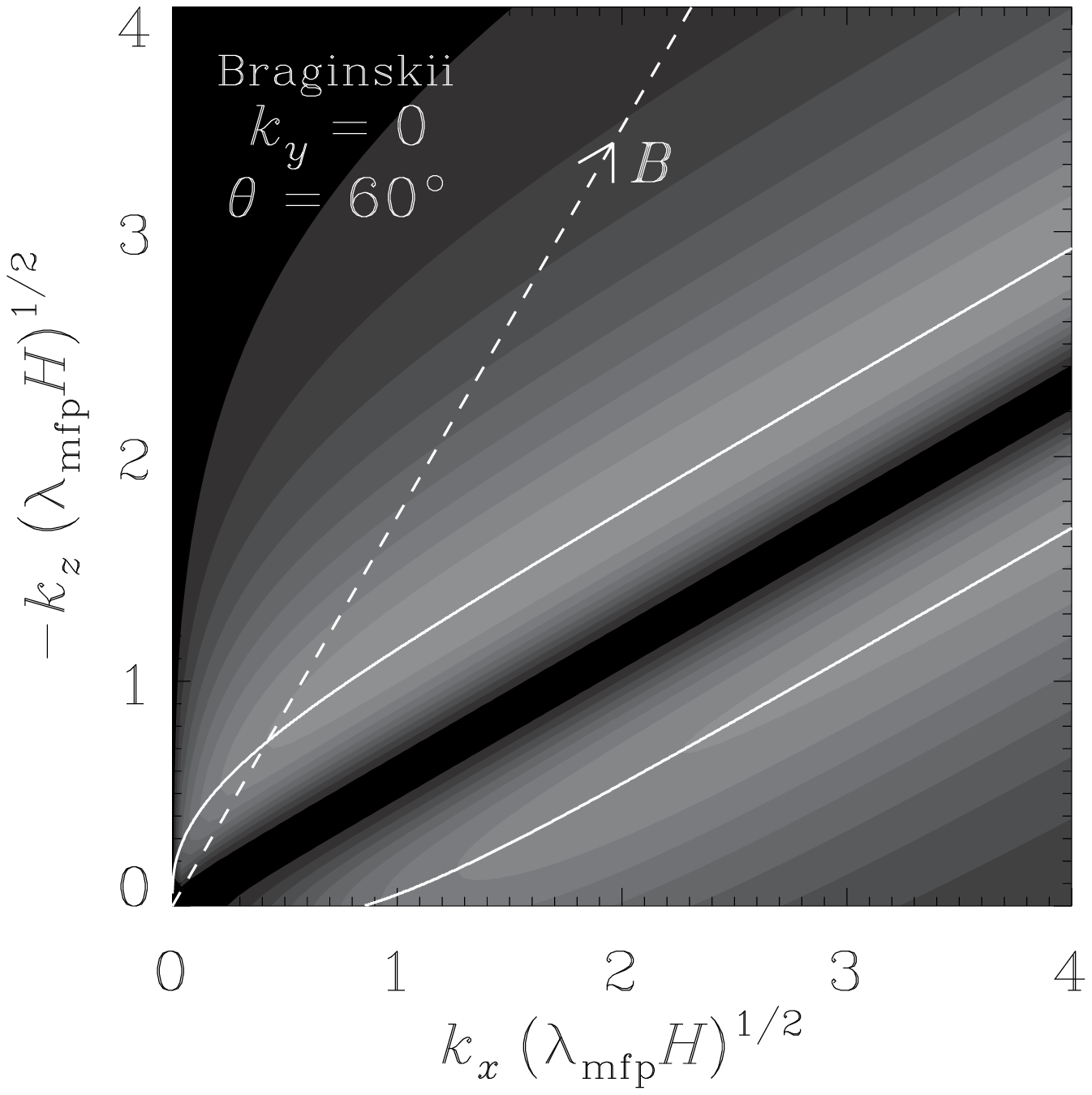}\quad
\includegraphics[height=2.2in]{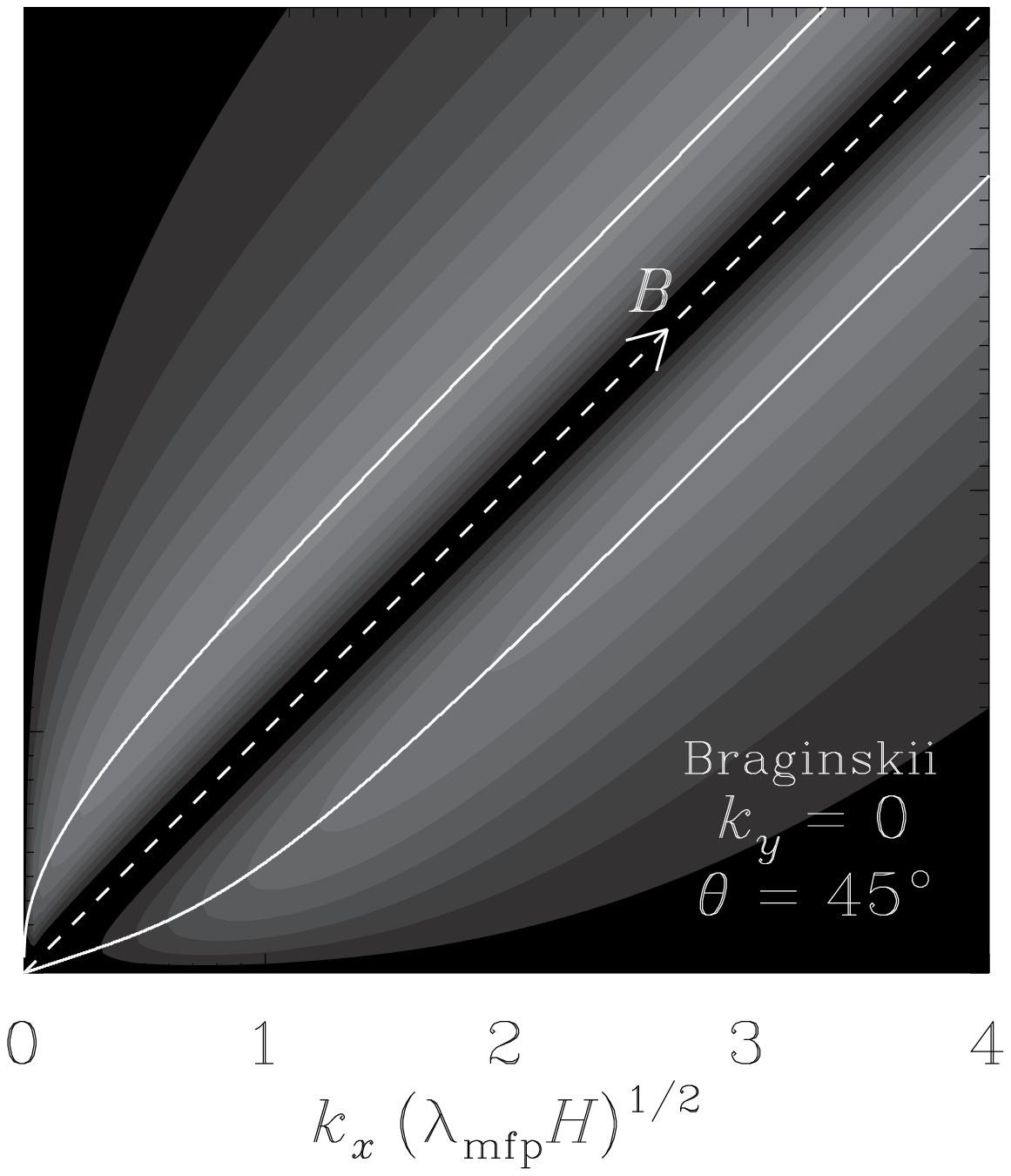}\quad
\includegraphics[height=2.2in]{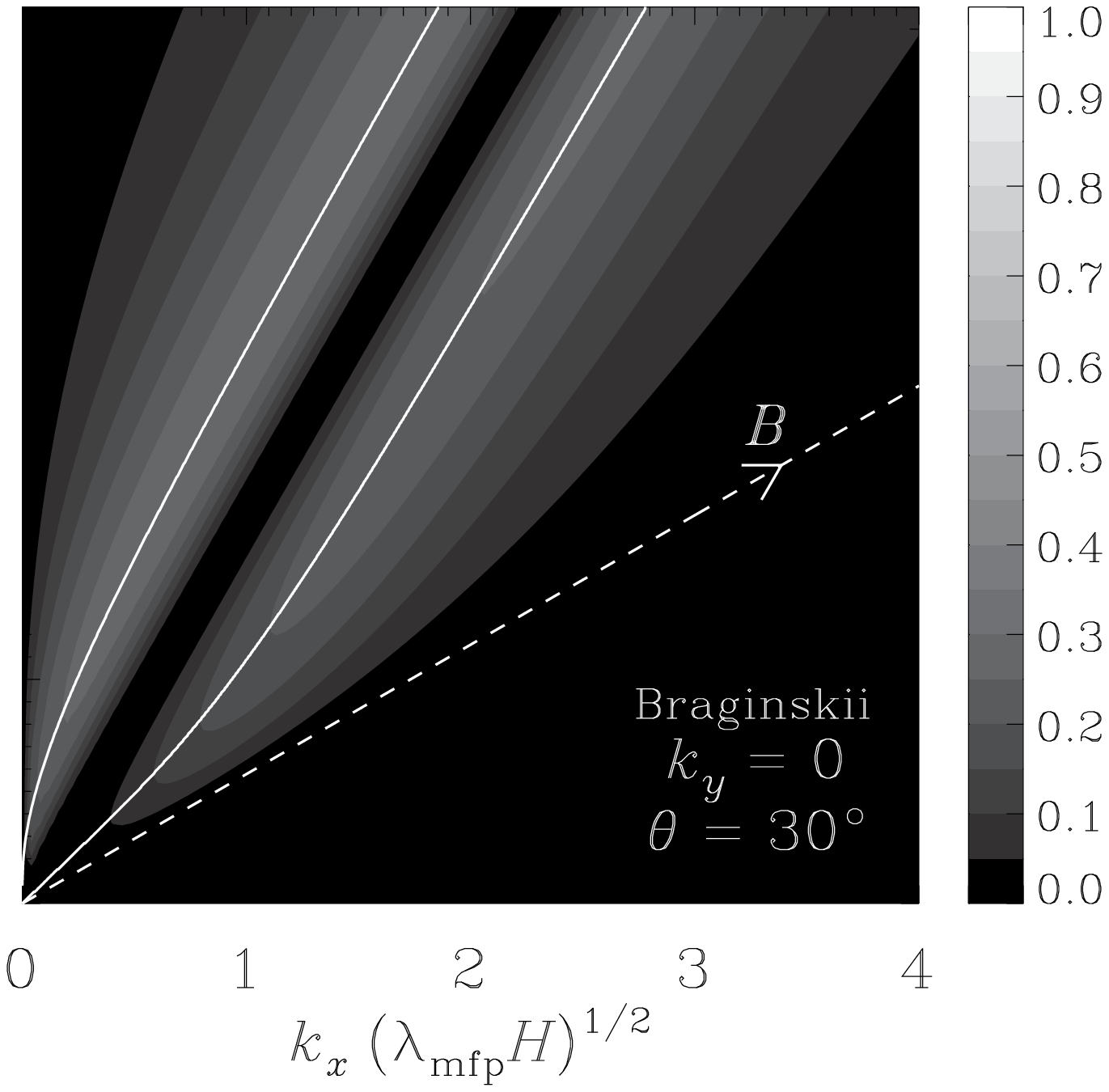}
\newline
\caption{HBI growth rate (normalised to $\sqrt{g \, {\rm d} \ln T / {\rm d} z}$) for $k_y=0$, ${\rm d} \ln T / {\rm d} \ln p=-1$, and various magnetic field orientations $\theta \equiv \cos^{-1} \left( b_x  \right)$. Magnetic tension is neglected; its effect is discussed in Section \ref{sec:hbitension}. Braginskii viscosity is included in the bottom row of plots. Each contour represents an increase in the growth rate by 5 per cent. The dashed line denotes the direction of the background magnetic field. The solid lines (given asymptotically by eq. \ref{eqn:linefit} with Braginskii viscosity and eq. \ref{eqn:approxhbi} without Braginskii viscosity) trace the maximum growth rate for a given total wavenumber $k$; the maximum growth rate is given by eq. \ref{eqn:approxgmax} with Braginskii viscosity and asymptotically at $k_{||} (\mfp H)^{1/2} \gg 1$ by eq. \ref{eqn:hbimax} without Braginskii viscosity. The overall maximum growth rates in each of the Braginskii-HBI plots (bottom row) are reduced by a factor $\simeq$$1.66$ relative to those in the respective standard HBI plots (top row). The thick black diagonal region in each plot is where $k_{||} = 0$.}
\label{fig:hbi_kxkz_angles}
\end{figure*}

Despite all this, there {\em are} modes that remain unstable to the HBI and retain non-negligible growth rates. However, it turns out that they are confined to a thin band of wavenumber space in which conduction is fast but viscous damping is small:
\setcounter{eqold}{\value{equation}}\setcounter{eqbid}{0}\addtocounter{eqold}{1}
\renewcommand{\theequation}{\arabic{eqold}\alph{eqbid}}\addtocounter{eqbid}{1}
\begin{equation}
\cond \gtrsim \dyn \gtrsim \visc ,
\end{equation}
or, using the definitions (\ref{eqn:wcond}) and (\ref{eqn:wvisc}),
\addtocounter{eqbid}{1}
\begin{equation}
3 \, k_{||} ( \mfp H )^{1/2} \; \gtrsim \; 1 \; \gtrsim \; k_{||} ( \mfp H )^{1/2} .
\end{equation}
\renewcommand{\theequation}{\arabic{equation}}\setcounter{equation}{\value{eqold}}
Using the fact that $k_\perp \simeq k$ for the fastest-growing Braginskii-HBI modes, it is possible to obtain analytic solutions for the maximum growth rate and fastest growing wavenumber. Defining $\varepsilon \equiv \visc / \cond \sim 0.1$, the maximum growth rate
\begin{eqnarray}\label{eqn:approxgmax}
\lefteqn
{
\sigma_{\rm max} = \frac{\sigma_{\rm HBI,max}}{( 1 - \varepsilon )} \left[ 1 - 2 \varepsilon^{1/2} \left( 2 + \frac{2}{5} \left| \D{\ln T}{\ln p} \right| \right)^{1/2} \right.
}
\nonumber\\*&&
\left. \mbox{} \times \left( 1 + \varepsilon + \frac{2 \varepsilon}{5} \left| \D{\ln T}{\ln p} \right| \right)^{1/2} + \varepsilon \left( 3 + \frac{4}{5} \left| \D{\ln T}{\ln p} \right| \right) \right]^{1/2}
\nonumber\\*
\end{eqnarray}
occurs at a parallel wavenumber satisfying 
\begin{eqnarray}\label{eqn:kmax}
k^2_{||} \mfp H & = & \frac{\sigma^2_{\rm HBI,max} - \sigma^2_{\rm max} ( 1 + \varepsilon ) }{3 \, \sigma_{\rm max} \, \dyn}
\\*
& \approx & \varepsilon^{1/2} \, \frac{2b_z}{3} \left( \frac{2}{5} + 2 \left| \D{\ln p}{\ln T} \right | \right)^{1/2} + \mc{O} ( \varepsilon ) .
\label{eqn:approxkmax}
\end{eqnarray}
Since $k_{||} = k_x b_x + k_z b_z$, equation (\ref{eqn:approxkmax}) implies that the maximum growth rate is attained along two straight lines in the $( k_x , k_z )$ plane given by
\begin{eqnarray}\label{eqn:linefit}
\lefteqn
{
k_z (\mfp H)^{1/2} \approx - \frac{b_x}{b_z} \, k_x (\mfp H)^{1/2} }
\nonumber \\*&& \mbox{} 
\pm \varepsilon^{1/4} \left( \frac{2}{3 b_z} \right)^{1/2}  \left( \frac{2}{5} + 2 \left| \D{\ln p}{\ln T} \right| \right)^{1/4} + \mc{O} ( \varepsilon^{3/4} ).
\end{eqnarray}
This behaviour can be seen in Fig. \ref{fig:hbi_kxkz_angles}, which exhibits HBI growth rates in the $( k_x , k_z )$ plane with (upper row) and without (lower row) Braginskii viscosity for $k_y = 0$ and various magnetic field orientations. The solid lines trace the maximum growth rate through wavenumber space; they quickly asymptote to equation (\ref{eqn:approxhbi}) without Braginskii viscosity and equation (\ref{eqn:linefit}) with Braginskii viscosity.

For a fiducial cool-core temperature profile ${\rm d} \ln T / {\rm d} \ln p = -1$, equations (\ref{eqn:approxgmax}) and (\ref{eqn:kmax}) give $\sigma_{\rm max} = 0.57 b_z \, \dyn$ and $k_{||} ( \mfp H )^{1/2} = 0.60 b^{1/2}_z$, respectively. With typical values of $H / \mfp \sim 10^2$--$10^3$ in the inner $\sim$$200~{\rm kpc}$ of cool-core clusters where the temperature increases with height, this implies $k_{||} H \sim 6 b^{1/2}_z $--$19 b^{1/2}_z$ (increasing inwards). {\em These modes are quite extended along the magnetic field direction and cannot be considered local.} This is likely to have important implications for the non-linear evolution of the HBI, particularly as the HBI reorients the mean magnetic field to be more and more perpendicular to the temperature gradient. For example, taking ${\rm d} \ln T / {\rm d} \ln p = -1$ and $H / \mfp = 200$, the parallel wavelength of maximum growth $\lambda_{||{\rm ,max}}$ is equal to the thermal-pressure scale-height $H$ when the magnetic field makes an angle of $\theta \simeq 33^\circ$ with respect to the $x$-axis. Thus, the field-line insulation found by many numerical simulations to be a consequence of the standard HBI \citep[e.g.][]{pqs09,brbp09} may not be as complete as is currently believed.

Note further that equation (\ref{eqn:approxkmax}) in the limit $\varepsilon \rightarrow 0$ does {\em not} reduce to the no-Braginskii case (eq. \ref{eqn:approxhbi}). Moreover, the relationship between $k_{||}$ and $k_\perp$ for the fastest growing modes discontinuously changes from $k_{||} \propto k^{1/2}_\perp$ without Braginskii viscosity (eq. \ref{eqn:approxhbi}) to $k_{||} \sim {\rm const}$ with Braginskii viscosity (eq. \ref{eqn:approxkmax}) for perpendicular wavenumbers satisfying
\begin{equation}
k_\perp \left( \mfp H \right)^{1/2} \gtrsim 0.5 \left| \D{\ln p}{\ln T} \right|^{1/4} \frac{1}{b^{1/2}_z} .
\end{equation}
This reflects the fact that including fast anisotropic heat conduction while neglecting Braginskii viscosity is a singular limit of the equations.

\subsubsection{Case of $b_x k_y \ne 0$: Alfv\'{e}nic HBI}\label{sec:alfvenichbi}

If $b_x k_y \ne 0$, the situation is actually worse:
\begin{eqnarray}\label{eqn:alfvenhbi}
\lefteqn
{
\sigma \simeq {\rm i} \left( g \D{z}{\ln T} \frac{b^2_x k^2_y}{k^2_\perp} \right)^{1/2} + \frac{\dyn^2}{\visc} \left| \D{\ln p}{\ln T} \right| \frac{k^2}{2k^2_\perp} \left( \frac{b^2_x k^2_y}{k^2_\perp} - \frac{\mc{K}}{k^2} \right)
}
\nonumber\\*
\end{eqnarray}
to leading order in $\dyn / \visc \ll 1$. The HBI becomes a slowly-growing overstability for wavevectors satisfying
\begin{equation}
k^2_y > \left| \frac{b_x}{b_z} k_{||} ( \bb{k} \btimes \eb )_y \right| - ( \bb{k} \btimes \eb )^2_y ,
\end{equation}
a weakly-damped oscillation for wavevectors satisfying
\begin{equation}
k^2_y < \left| \frac{b_x}{b_z} k_{||} ( \bb{k} \btimes \eb )_y \right| - ( \bb{k} \btimes \eb )^2_y ,
\end{equation}
and a pure oscillation if the left- and right-hand sides of these inequalities are in fact equal. Indeed, in the limit $\dyn / \visc \ll 1$ equations (\ref{eqn:eigenrho})--(\ref{eqn:DT}) imply
\begin{equation}
\frac{\delta \rho}{\rho} \simeq \xi_z \D{z}{\ln T}  + {\rm i} \, \mc{O} \left( \frac{\dyn}{\visc} \right) + \mc{O} \left( \frac{\dyn^2}{\visc^2} \right) ,
\end{equation}
\begin{equation}
\frac{\Delta T}{T} \sim {\rm i} \, \mc{O} \left( \frac{\dyn}{\visc} \right) + \mc{O} \left( \frac{\dyn^2}{\visc^2} \right) ,
\end{equation}
\begin{equation}
\frac{\delta B_{||}}{B} \sim {\rm i} \, \mc{O} \left( \frac{\dyn}{\visc} \right) + \mc{O} \left( \frac{\dyn^2}{\visc^2} \right) .
\end{equation}
By effectively reorienting magnetic field perturbations to be nearly perpendicular to the background magnetic field, Braginskii viscosity prevents slow-mode perturbations from tapping into the free energy carried by the background heat flux. To highest order in $\dyn / \visc$, these modes appear as HBI-stable Alfv\'{e}n waves whose magnetic tension has been effectively increased by the adverse temperature gradient.

\subsubsection{Effect of magnetic tension on the HBI}\label{sec:hbitension}

When both Braginskii viscosity and magnetic tension are included there are two parallel-wavenumber cutoffs, the relative magnitude of which may play an important role in the evolution and non-linear saturation of the HBI. Roughly speaking, magnetic tension is appreciable for parallel wavenumbers satisfying $k_{||{\rm ,max}} v_{\rm A} \gtrsim \sigma_{\rm max}$, where $k_{||{\rm ,max}}$ and $\sigma_{\rm max}$ are given by equations (\ref{eqn:kmax}) and (\ref{eqn:approxgmax}) respectively. For a fiducial cool-core temperature profile, ${\rm d} \ln T / {\rm d} \ln p = -1$, this amounts to an upper limit on the plasma beta parameter of $b_z \beta \lesssim H / \mfp$. For $\beta$ less than this value, magnetic tension -- not Braginskii viscosity -- sets the fastest-growing mode.

\subsection{${\rm d} T / {\rm d} z < 0$: Magnetothermal Instability}\label{sec:mti}

Next we investigate the effects of Braginskii viscosity on the stability of a stratified atmosphere in which the temperature increases in the direction of gravity, i.e. ${\rm d} T / {\rm d} z < 0$. Such an atmosphere was shown by \citet{balbus00} to be susceptible to the MTI if $\mc{K} > 0$.

%
%
\begin{figure}
\centering
\includegraphics[width=2.7in]{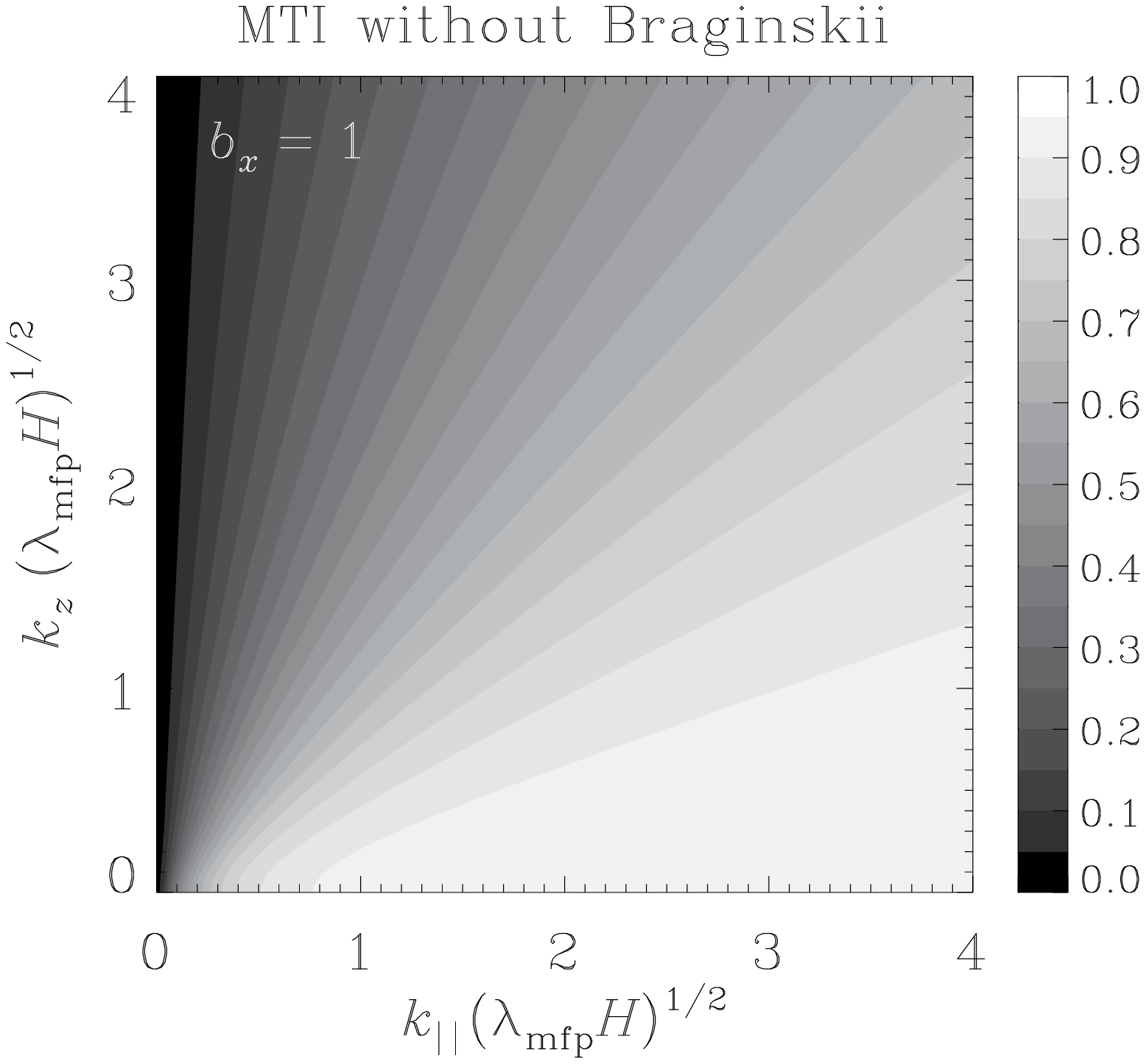}
\newline
\newline
\includegraphics[width=2.7in]{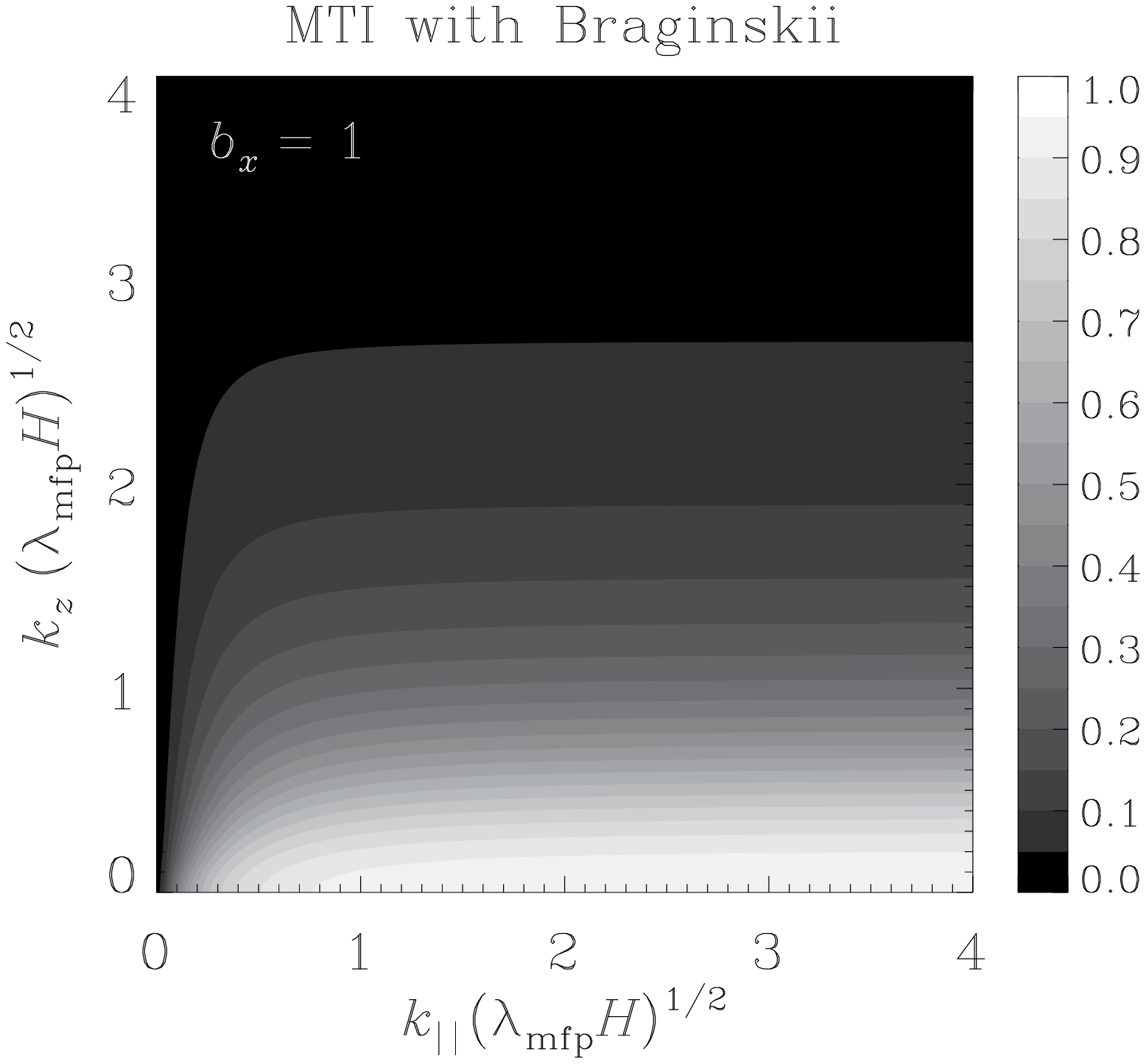}
\newline
\caption{MTI growth rate (normalised to the maximum growth rate $\sqrt{ -g \, {\rm d} \ln T / {\rm d} z}$) for $k_y = 0$ without ({\it top}) and with ({\it bottom}) Braginskii viscosity for a stratified thermal layer with ${\rm d} \ln T / {\rm d} \ln p = 1/3$ threaded by a horizontal magnetic field ($b_x = 1$). Magnetic tension is neglected; its effect is discussed in Section \ref{sec:mtitension}. Each contour represents an increase in the growth rate by 5 per cent. Braginskii viscosity suppresses MTI growth rates for $k \ne k_{||}$ (see eqs \ref{eqn:mtikperp} and \ref{eqn:narrowband}). Galaxy clusters typically have $H / \mfp \sim 10$--$100$ at radii for which the temperature profile decreases outwards, and so the maximum wavenumbers for each axis span the range $k H \sim 13$--$40$.}
\label{fig:mti_kxkz_bx}
\end{figure}

%
%
\begin{figure*}
\centering
\includegraphics[height=2.2in]{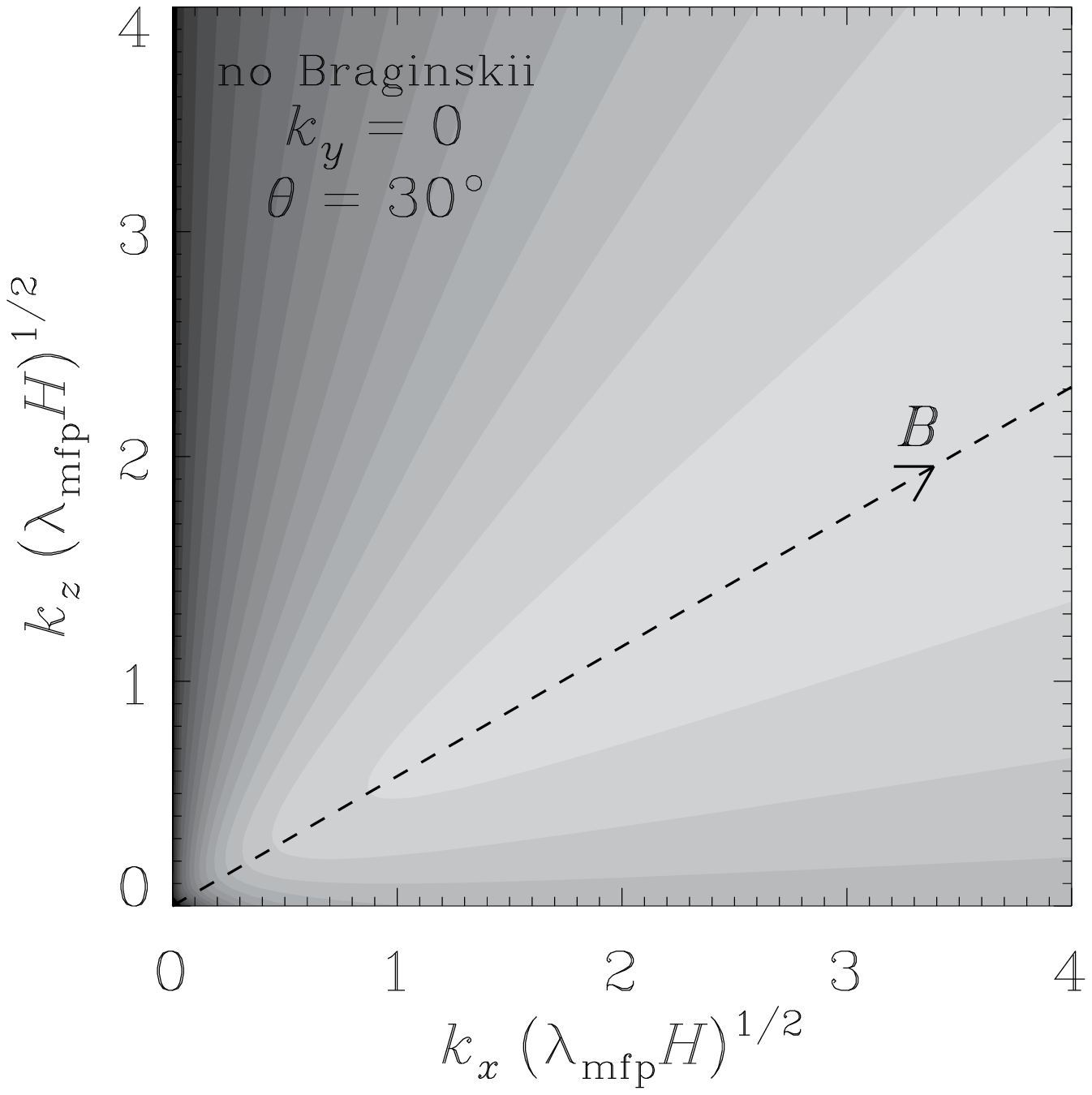}\quad
\includegraphics[height=2.2in]{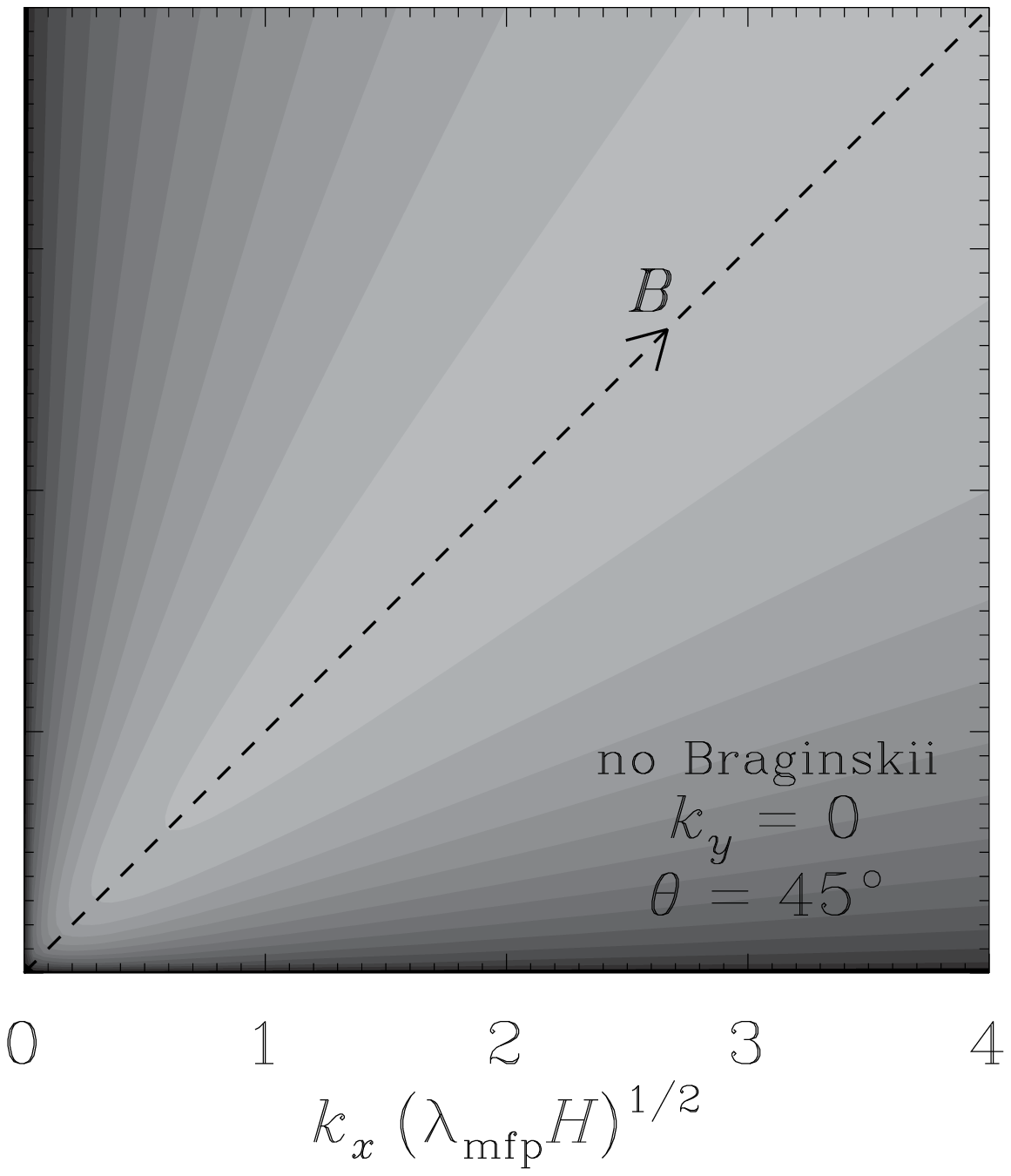}\quad
\includegraphics[height=2.2in]{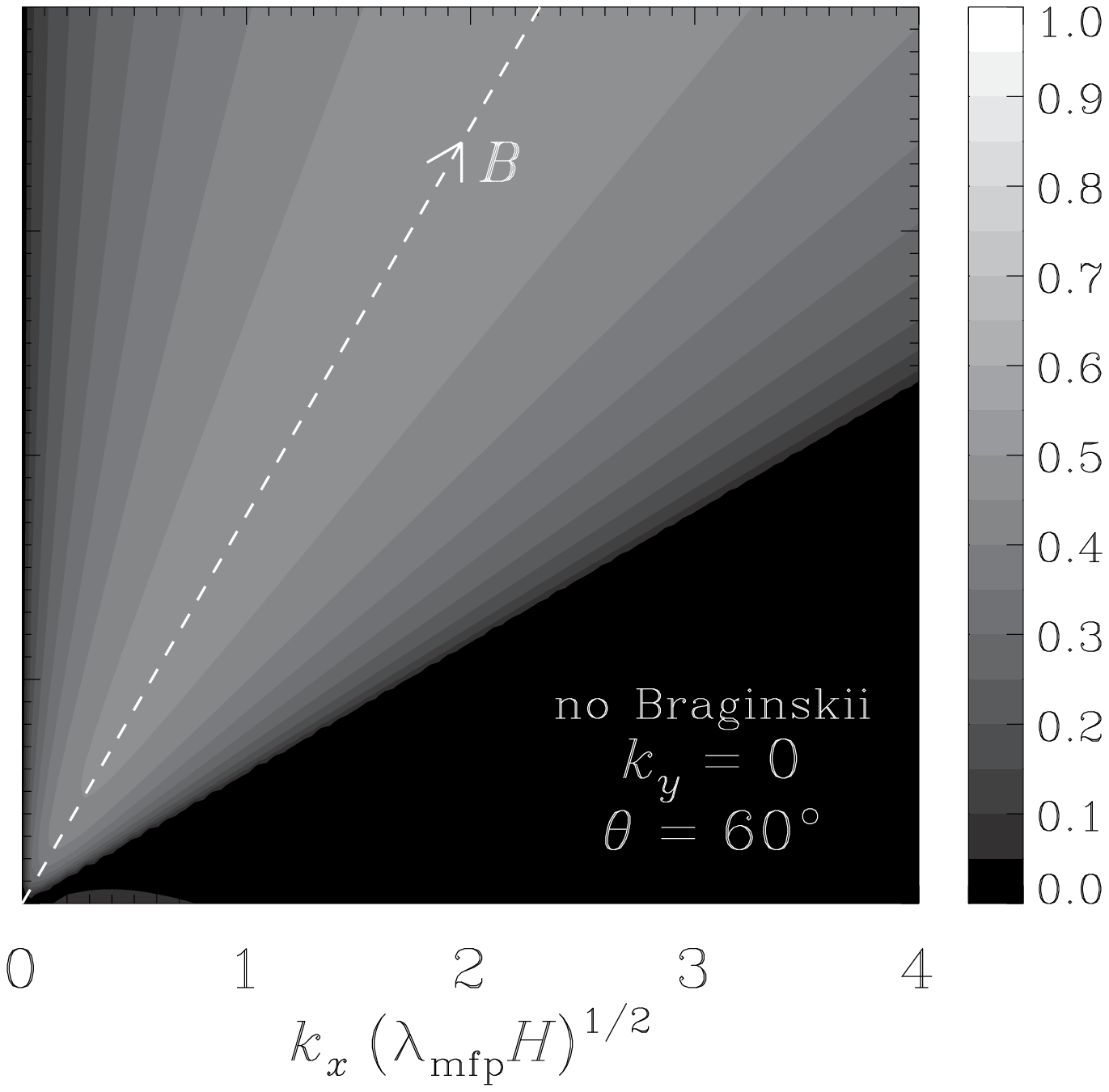}
\newline
\newline
\includegraphics[height=2.2in]{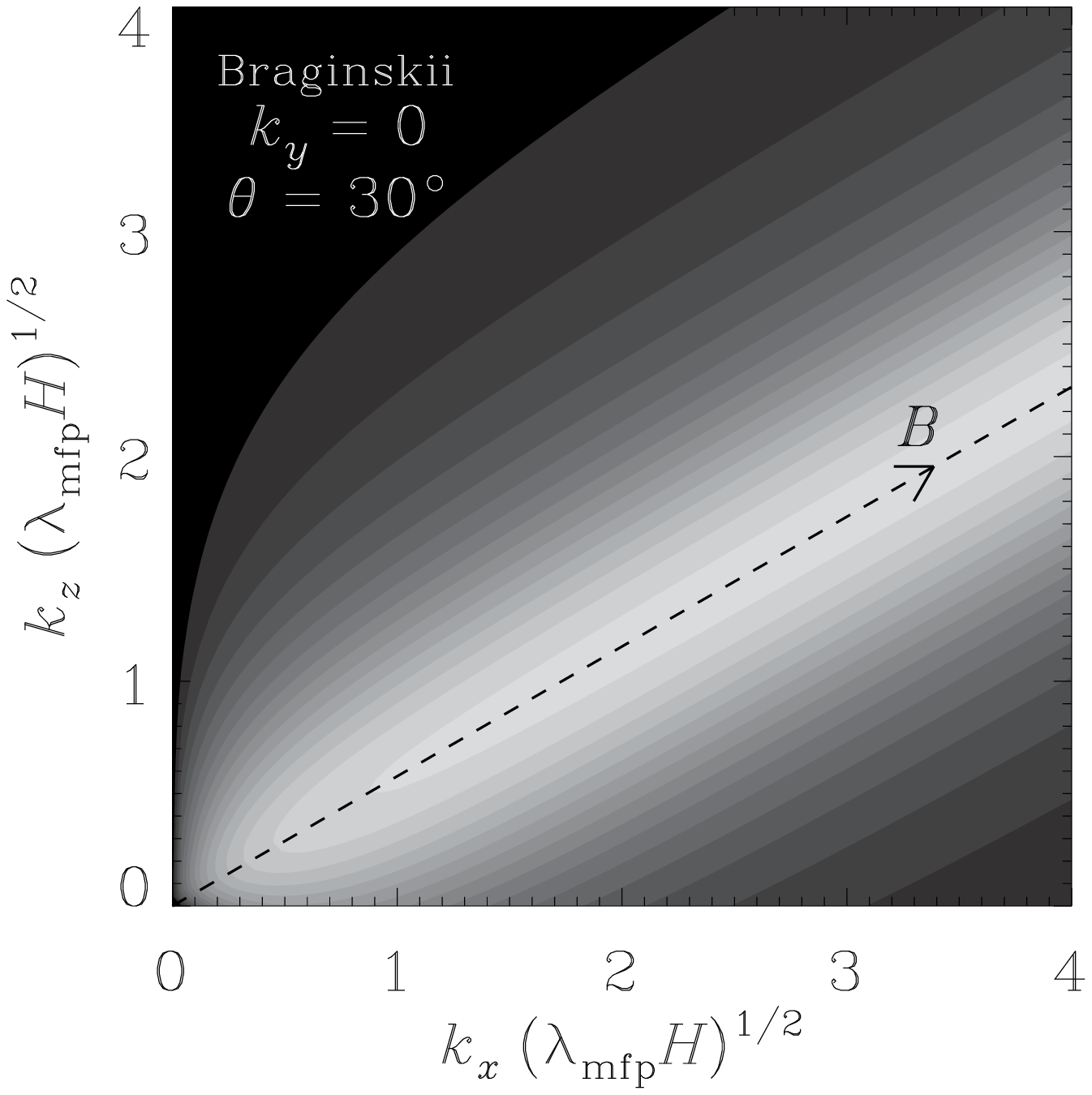}\quad
\includegraphics[height=2.2in]{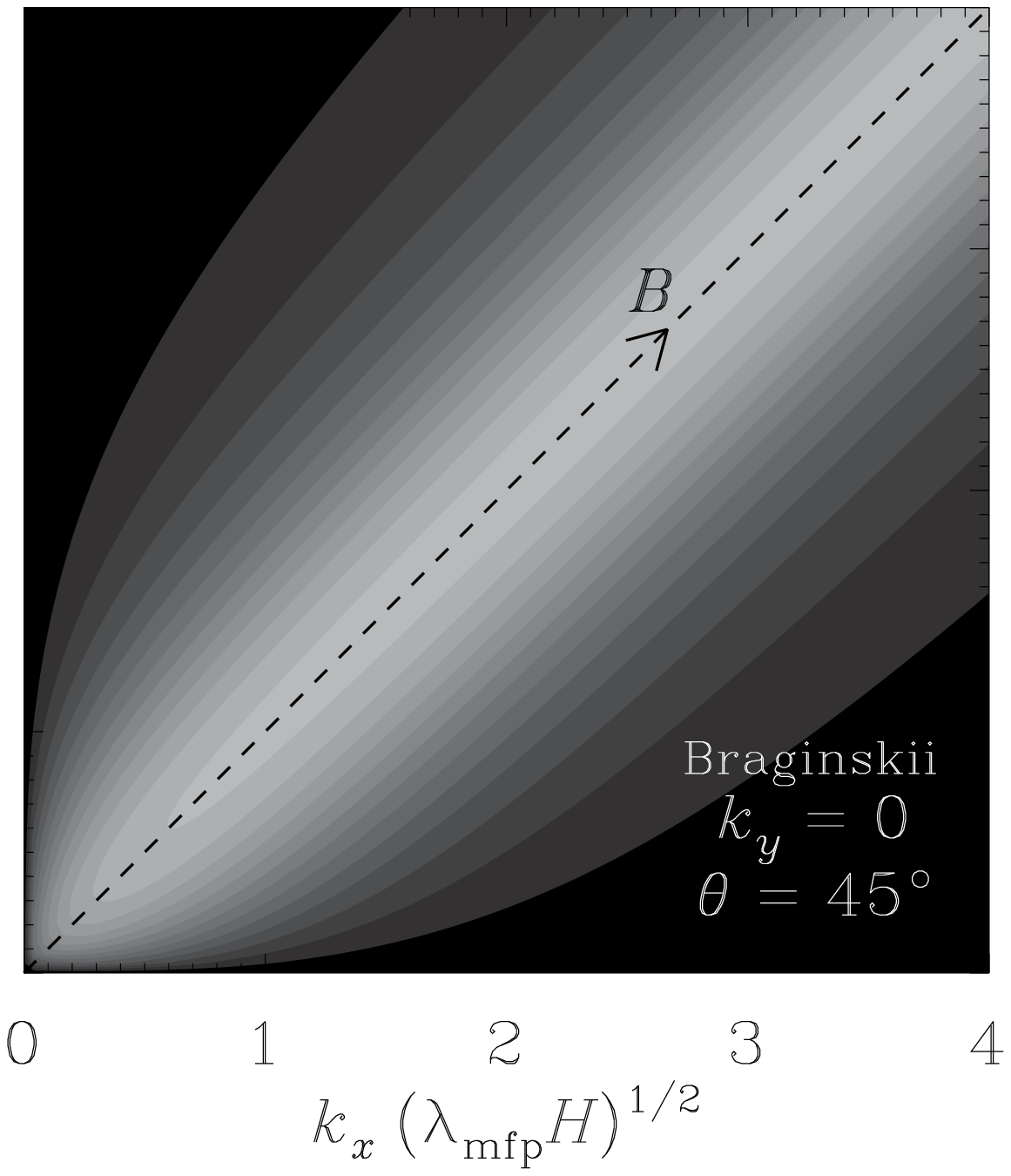}\quad
\includegraphics[height=2.2in]{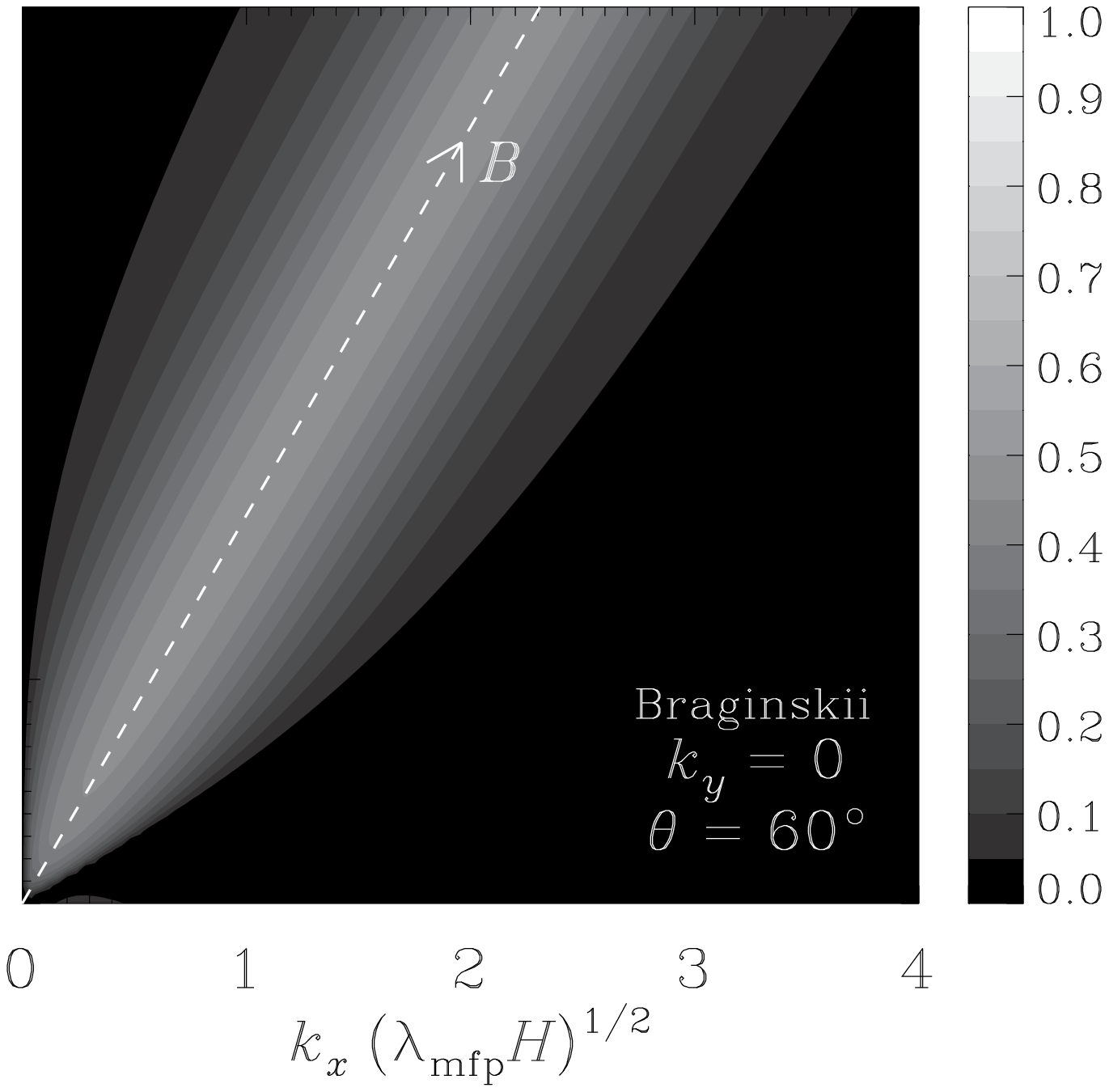}
\newline
\caption{MTI growth rate (normalised to $\sqrt{-g \, {\rm d} \ln T / {\rm d} z}$) for $k_y=0$, ${\rm d} \ln T / {\rm d} \ln p = 1/3$, and various magnetic field orientations $\theta \equiv \cos^{-1} \left( b_x \right)$. Magnetic tension is neglected; its effect is discussed in Section \ref{sec:mtitension}. Braginskii viscosity is included in the bottom row of plots. Each contour represents an increase in the growth rate by 5 per cent. The dashed line denotes the direction of the background magnetic field, which also traces the maximum growth rate for a given total wavenumber $k$. The maximum growth rate is given by equation (\ref{eqn:mtimax}) and occurs for $k_\perp = 0$. Braginskii viscosity suppresses MTI growth rates for $k \ne k_{||}$ (see eqs \ref{eqn:mtikperp} and \ref{eqn:narrowband}).}
\label{fig:mti_kxkz_angles}
\end{figure*}

\subsubsection{Case of $k_y = 0$: Standard MTI with and without Braginskii viscosity}\label{sec:standardmti}

Consider first the case $k_y = 0$. Equation (\ref{eqn:disprel2}) shows that the maximum MTI growth rate
\begin{equation}\label{eqn:mtimax}
\sigma^2_{\rm MTI,max} = g \left| \D{z}{\ln T} \right| b^2_x
\end{equation}
occurs for $k_\perp = 0$, where we have assumed that $k_{||} H \ll b_x \beta^{1/2}$ (i.e. magnetic tension is negligible on the scales of interest). The physical reasons for this are simple. Since $\bb{k} \bcdot \delta \bb{B} = 0$, taking $k_\perp = 0$ implies $\delta B_{||} = 0$. This not only ensures that any background heat flux is unable to cool (heat) upwardly (downwardly) displaced fluid elements (see eq. \ref{eqn:DT}), but also that pressure anisotropy cannot damp these modes.

If we allow for a small wavenumber component perpendicular to the background field, $k^2_\perp \ll k^2$, the leading-order solution for the growth rate is given by
\begin{equation}\label{eqn:mtikperp}
\sigma^2 \simeq \sigma^2_{\rm MTI,max} \left(1 - \frac{k^2_\perp}{b^2_x k^2} \right) - \sigma_{\rm MTI,max} \, \visc \frac{k^2_\perp}{k^2} .
\end{equation}
The first negative contribution on the right-hand side of this equation is tied to the fact that generating a $\delta B_{||}$ implies $\bb{k} \bcdot \delta \eb \ne 0$. Having $\bb{k} \bcdot \delta \eb \ne 0$ causes upwardly displaced fluid elements, which are trying to heat and rise by the MTI, to be cooled by the locally divergent background heat flux (and vice versa for downward displacements). This reduces the efficiency of the MTI and implies that unstable modes with large growth rates are confined to a wedge in wavenumber space of width
\begin{equation}\label{eqn:mtiband}
k_\perp \lesssim b_x k_{||} .
\end{equation}
As $k_{||}$ increases, more and more perpendicular wavenumber space becomes available for fast-growing modes (see top panel of Fig. \ref{fig:mti_kxkz_bx}). 

When $\visc \gg \dyn$, however, this term is relatively unimportant when compared to the last term on the right-hand side of equation (\ref{eqn:mtikperp}). Modes with $k^2_\perp \ne 0$ are rapidly damped by Braginskii viscosity. This behaviour continues beyond the small values of $k_\perp$, where the growth rate then becomes
\begin{equation}\label{eqn:approxbragmti}
\sigma \simeq \frac{g}{\visc} \left| \D{z}{\ln T} \right| \left( \frac{b^2_x k^2}{k^2_\perp} - 1 \right)
\end{equation}
to leading order in $\dyn / \visc \ll 1$. These effects are evident in Figs \ref{fig:mti_kxkz_bx} and \ref{fig:mti_kxkz_angles}, which exhibit growth rates in the $( k_x , k_z )$ plane for various inclinations of the background magnetic field. From equation (\ref{eqn:approxbragmti}), we find that unstable modes with large growth rates are confined by Braginskii viscosity to a narrow band in wavenumber space of width
\begin{equation}\label{eqn:narrowband}
k_\perp \left( \mfp H \right)^{1/2} \lesssim \left( \D{\ln p}{\ln T} \right)^{1/4} b^{1/2}_x
\end{equation}
about the line $\bb{k} \, || \, \eb$. In contrast with the no-Braginskii case (eq. \ref{eqn:mtiband}), going to larger $k_{||}$ does {\em not} open up more perpendicular wavenumber space for fast-growing modes. 

%
%
\begin{figure*}
\centering
\includegraphics[height=2.2in]{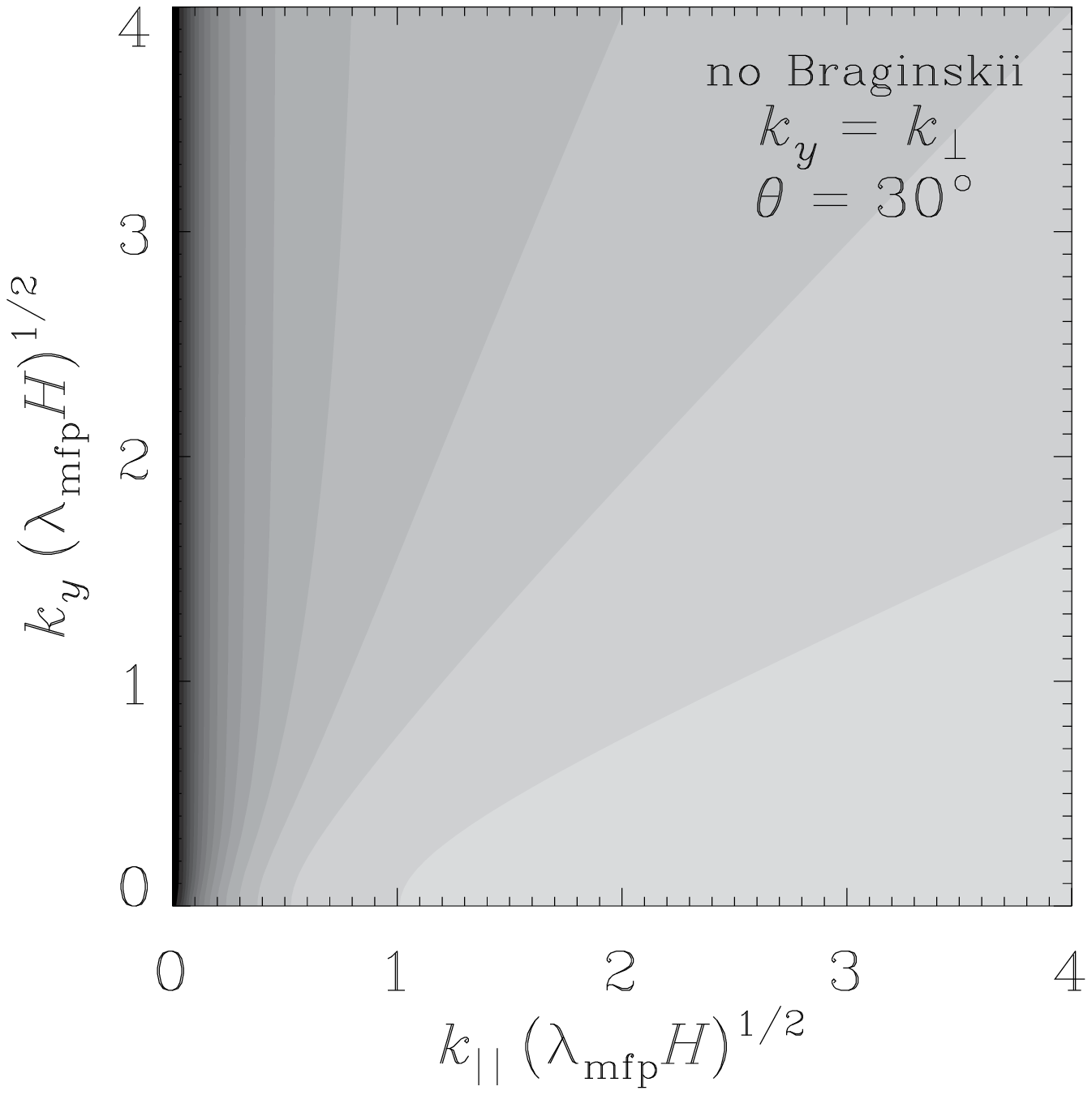}\quad
\includegraphics[height=2.2in]{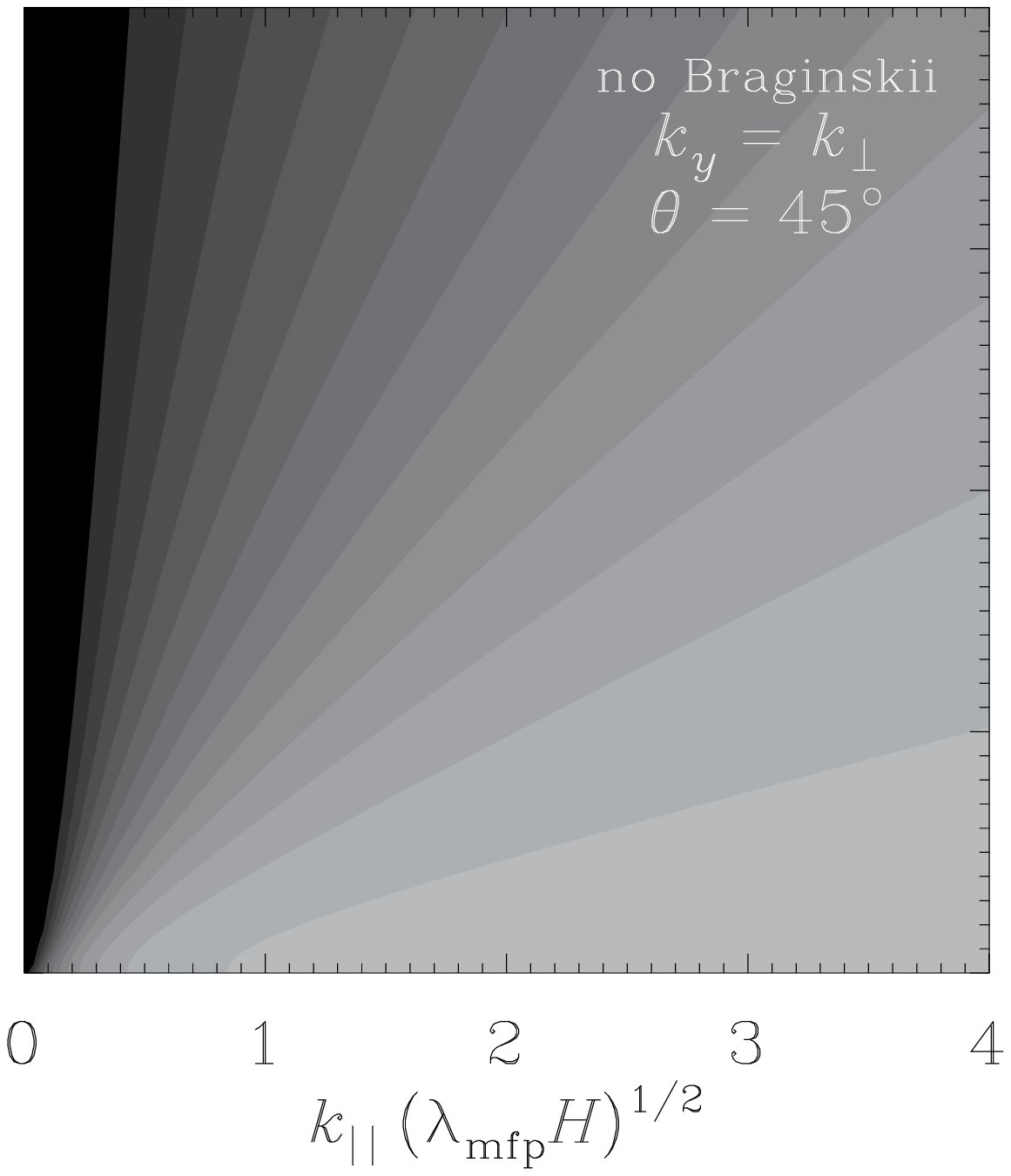}\quad
\includegraphics[height=2.2in]{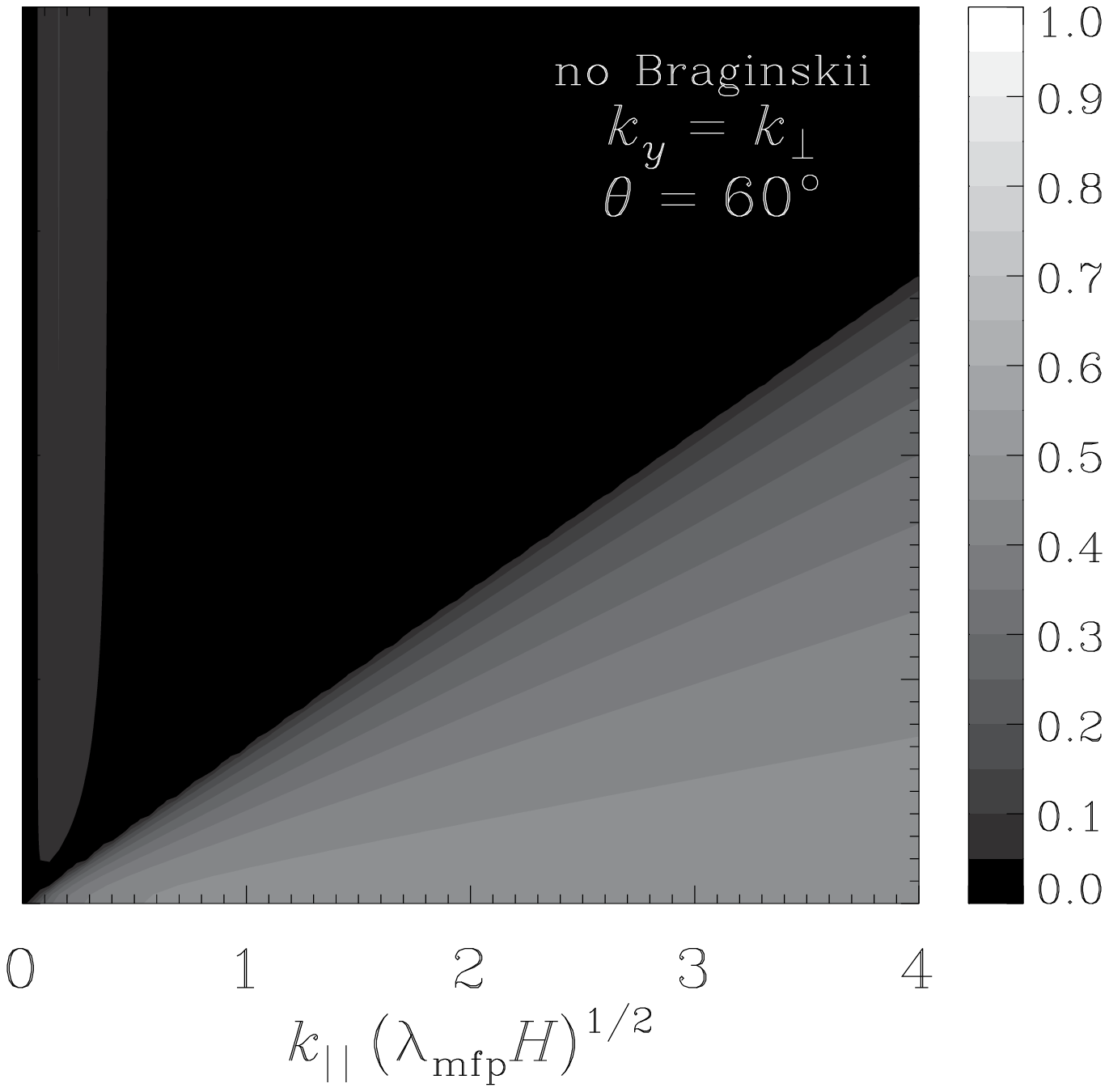}
\newline
\newline
\includegraphics[height=2.2in]{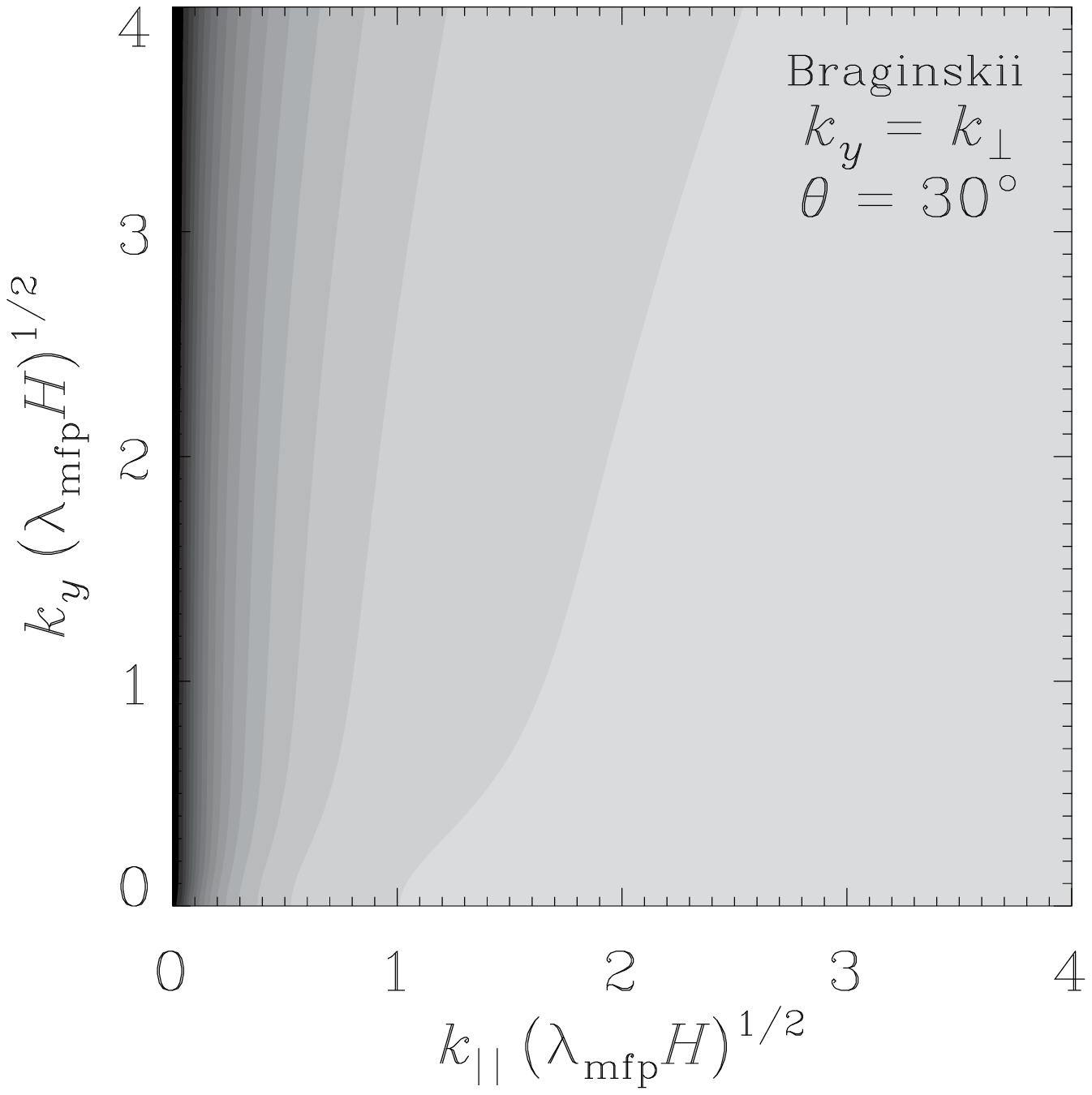}\quad
\includegraphics[height=2.2in]{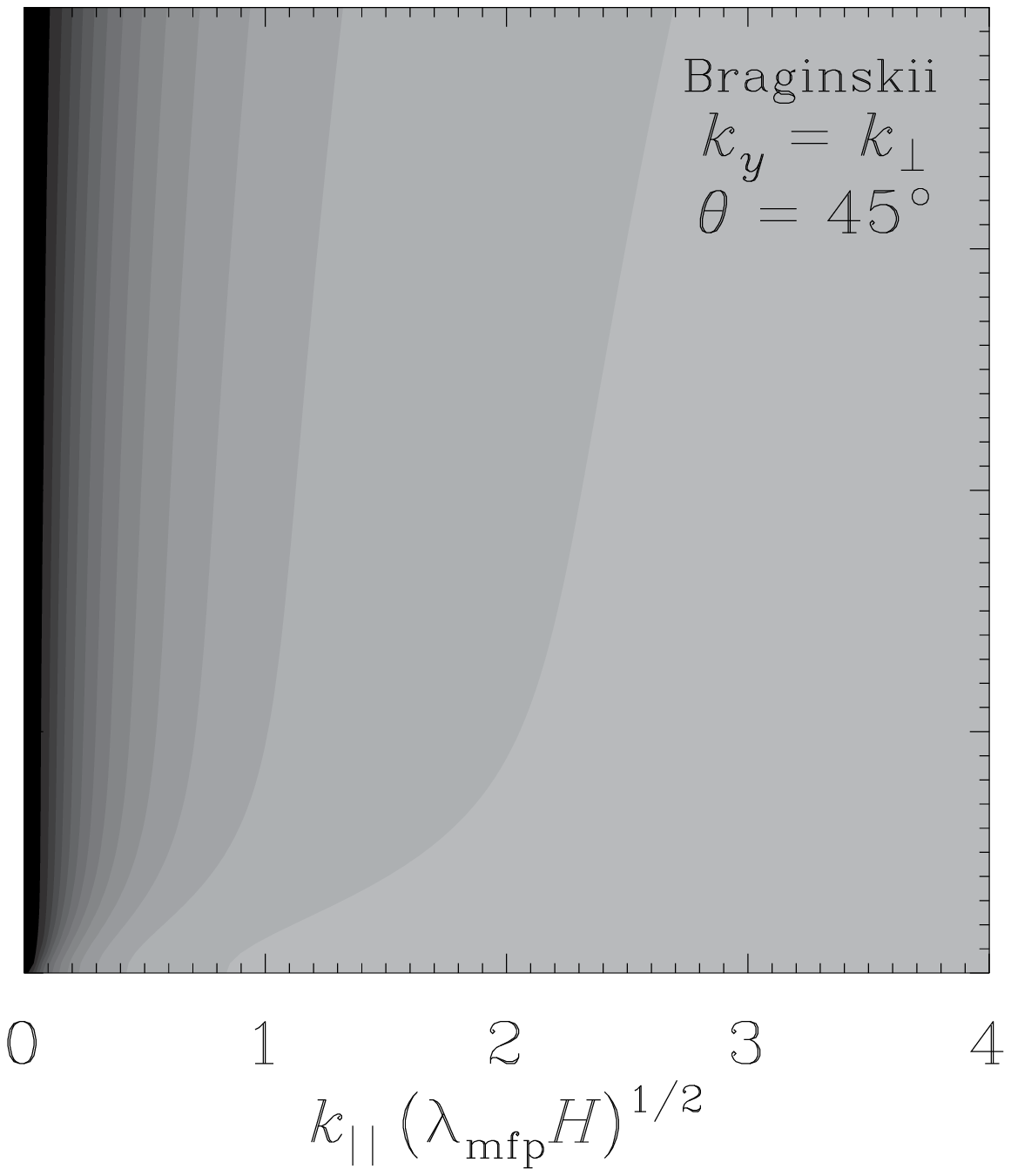}\quad
\includegraphics[height=2.2in]{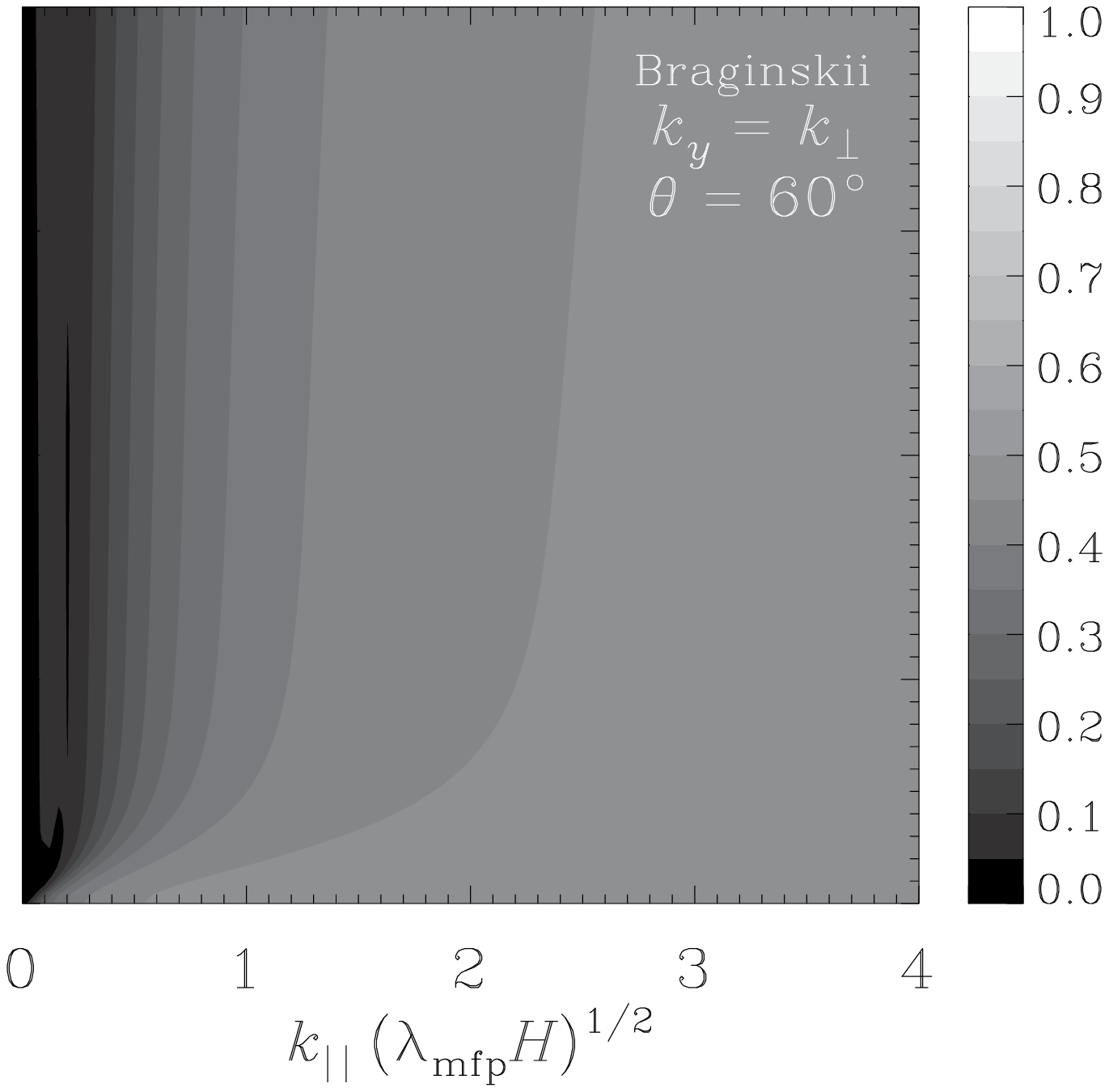}
\newline
\caption{MTI growth rate (normalised to $\sqrt{-g \, {\rm d} \ln T / {\rm d} z}$) for $k^2_y = k^2_\perp$, ${\rm d} \ln T / {\rm d} \ln p=1/3$, and various magnetic field orientations $\theta \equiv \cos^{-1} \left( b_x \right)$. Magnetic tension is neglected; Braginskii viscosity is included in the bottom row of plots. Each contour represents an increase in the growth rate by 5 per cent. Many $k_y \ne 0$ modes that are either stable or only grow slowly in the absence of Braginskii viscosity become unstable with growth rate $\sigma_{\rm MTI,max}$ when Braginskii viscosity is included.}
\label{fig:mti_kpky}
\end{figure*}

\subsubsection{Case of $b_x k_y \ne 0$: Alfv\'{e}nic MTI}\label{sec:alfvenicmti}

Recall equation (\ref{eqn:disprel2}):
\begin{eqnarray}\label{eqn:disprel3}
\lefteqn
{
\widetilde{\sigma}^2 \left( \widetilde{\sigma}^2 + \sigma \visc \frac{k^2_\perp}{k^2} + g \D{z}{\ln T} \frac{\mc{K}}{k^2} \right) \simeq - \sigma \visc \, g \D{z}{\ln T} \frac{b^2_x k^2_y}{k^2} ,
}
\nonumber\\*
\end{eqnarray}
which is the general dispersion relation (\ref{eqn:disprel1}) written in the fast-conduction limit ($\cond \gg \dyn \sim \sigma$). When $b^2_x k^2_y \ne 0$, the right-hand side of this equation becomes active and leads to behaviour otherwise absent without Braginskii viscosity. The Alfv\'{e}n-mode branch of the dispersion relation is now coupled to the slow-mode branch; slow-mode perturbations induce an Alfv\'{e}nic response.

Consider further the limit $\visc \gg \dyn$. Then we obtain eq. (\ref{eqn:brag2}):
\begin{equation}
\widetilde{\sigma}^2 \simeq g \left| \D{z}{\ln T} \right| \frac{b^2_x k^2_y}{k^2_\perp} .
\end{equation}
{\em This is always unstable, regardless of the sign of $\mc{K}$}, and is maximal when $k^2_y = k^2_\perp$ (i.e. the projection of $\bb{k}$ onto the $x$-$z$ plane is parallel to the magnetic field). In other words, wavevectors with large parallel components and {\em any} component along the $y$-axis grow at $\sigma_{\rm MTI,max}$ (see Fig. \ref{fig:mti_kpky}). Indeed, one can readily show from equation (\ref{eqn:disprel3}) that the growth rate for these $k^2_y = k^2_\perp$ Alfv\'{e}nic MTI modes is 
\begin{equation}
\sigma \simeq \sigma_{\rm MTI,max} - \frac{\omega^2_{\rm dyn}}{\visc}  \D{\ln p}{\ln T} \frac{b^2_z}{2} 
\end{equation}
to leading order in $\dyn / \visc$, so that one only requires
\begin{equation}\label{eqn:kpalfvenmti}
k_{||} ( \mfp H )^{1/2} \gtrsim \left( \D{\ln p}{\ln T} \right)^{1/4} \frac{b_z}{b^{1/2}_x} 
\end{equation}
to bring the growth rate close to $\sigma_{\rm MTI,max}$ (e.g. see the rightmost panel in the bottom row of Fig. \ref{fig:mti_kpky}). One consequence is that it is no longer necessary to go to very large parallel wavenumbers (and risk stabilisation by magnetic tension) just to marginally destabilise $k_y \ne 0$ modes at small $b_x$ (e.g. see the rightmost panel in the top row of Fig. \ref{fig:mti_kpky}).

The physical origin of this new behaviour may be uncovered by computing the eigenvectors (\ref{eqn:eigenrho}), (\ref{eqn:eigenb}) and (\ref{eqn:DT}) to leading order in $\dyn / \visc$:
\begin{equation}\label{eqn:alfmtirho}
\frac{\delta \rho}{\rho} \simeq - \xi_z \left| \D{z}{\ln T} \right| \left[ 1 - \frac{\dyn}{\visc} \left( \D{\ln p}{\ln T} \right)^{1/2} \frac{2b^2_z}{b_x} \right] ,
\end{equation}
\begin{equation}\label{eqn:alfmtib}
\frac{\delta B_{||}}{B} \simeq {\rm i} k_{||} \xi_z \, \frac{\dyn}{\visc} \left( \D{\ln p}{\ln T} \right)^{1/2} \frac{b_z}{b_x} .
\end{equation}
\begin{equation}\label{eqn:alfmtidt}
\frac{\Delta T}{T} \simeq - \xi_z \left| \D{z}{\ln T} \right| \frac{\dyn}{\visc} \left( \D{\ln p}{\ln T} \right)^{1/2} \frac{2 b^2_z}{b_x} .
\end{equation}
These are essentially MTI modes that have been freed from the unfavourable consequences of having $\bb{k} \bcdot \delta \eb \ne 0$ by rapid parallel viscous damping. Rapid Braginskii damping endows buoyantly-unstable slow-mode perturbations (eq. \ref{eqn:alfmtirho}) with Alfv\'{e}nic characteristics (i.e. perturbed magnetic fields and velocities that are predominantly oriented perpendicular to the background magnetic field -- eq. \ref{eqn:alfmtib}), which allows fluid elements to approximately maintain their temperature as they are displaced (eq. \ref{eqn:alfmtidt}). The presence of a non-zero $k_y$ is necessary in order to ensure $\delta B_{||}$ is vanishingly small for arbitrary $k_x$ and $k_z$, while simultaneously preserving the divergence-free constraint on the magnetic field. Therefore, many wavevectors for which $\mc{K} \le 0$ that are stable to the standard MTI (e.g. $k^2_y = k^2_\perp \geq k^2 b^2_x / b^2_z$, which is the black region in the upper-right panel of Fig. \ref{fig:mti_kpky}) are actually unstable with growth rates $\simeq$$\sigma_{\rm MTI,max}$. 

\subsubsection{Effect of magnetic tension on the MTI}\label{sec:mtitension}

When $k_y = 0$ the fastest growing MTI modes, those with $k = k_{||}$, are unaffected by Braginskii viscosity. In this case, magnetic tension provides the only parallel-wavenumber cutoff by suppressing wavenumbers for which $k_{||} v_{\rm A} \gtrsim \sigma_{\rm MTI, max}$, or
\begin{equation}
k H = k_{||} H \gtrsim \beta^{1/2} \left( \D{\ln p}{\ln T} \right)^{1/2} b_x .
\end{equation}
Setting $k H \sim 2 \pi$ provides a strict lower limit on beta below which local $k_y = 0$ MTI modes are stabilised:
\begin{equation}
\beta \lesssim \D{\ln T}{\ln p} \left( \frac{2 \pi}{b_x} \right)^2 .
\end{equation}
For a fiducial cluster temperature profile beyond the cooling radius, ${\rm d} \ln T / {\rm d} \ln p = 1/3$, this gives $\beta \lesssim 120 \, b^{-2}_x$.

If we relax our restriction on $k_y$, Braginskii viscosity drives Alfv\'{e}nic modes unstable by coupling them to buoyantly-unstable, rapidly-damped slow modes. Setting $k_{||} v_{\rm A} \sim \sigma_{\rm MTI,max}$ in equation (\ref{eqn:kpalfvenmti}) we find that, unless
\begin{equation}
\beta \lesssim \frac{H}{\mfp} \left( \D{\ln T}{\ln p} \right)^{1/2} \frac{b^2_z}{b^3_x} ,
\end{equation}
there are fast-growing Alfv\'{e}nic MTI modes. With typical values of $H / \mfp \sim 10$--$100$ and $\beta \sim 10^3$--$10^4$ in the outer regions of the ICM where the temperature decreases outwards, magnetic tension is unlikely to affect these modes except possibly when the field is nearly vertical ($b_x \ll 1$). Note that increasing $k_y$ does not increase magnetic tension since $\eb \perp \ey$.

\subsection{${\rm d} T / {\rm d} z < 0$: Heat-flux--driven Buoyancy Overstability}

The reader may have noticed a very thin vertical band of unstable modes in the $\theta = 60^\circ$ panels of Fig. \ref{fig:mti_kpky}. These modes, found recently by \citet{br10}, are $g$-modes driven overstable by a background heat flux. They become important only when the background magnetic field is vertical ($b_z = 1$, $\mc{K} = -k^2_\perp$), since this field orientation is (linearly) stable to the MTI. In this Section we analyse these modes while including Braginskii viscosity.\footnote{We have chosen not to present a similar analysis for the radiative-cooling--driven $g$-mode overstability also found by \citet{br10} for ${\rm d} T / {\rm d} z > 0$. The necessary assumption that the local cooling rate is comparable to the dynamical frequency conflicts with our assumption of an equilibrium background state that evolves slower than the instabilities do.}

We begin by finding the fastest-growing mode in the absence of Braginskii viscosity. Our task is greatly simplified by knowing {\it a priori} that $k_\perp \simeq k$ for these modes. We also neglect magnetic tension; we will verify {\em a posteriori} that it is unlikely to affect the fastest growing mode for conditions found in the outer regions of the ICM where this overstability may be present. Our dispersion relation (\ref{eqn:disprel1}) then becomes
\begin{equation}\label{eqn:disprel4}
\sigma^3 + \sigma^2 \cond + \sigma N^2 + \cond \, g \left| \D{z}{\ln T} \right| = 0 .
\end{equation}
Solutions of this equation have both real and imaginary parts, $\sigma = \gamma + {\rm i} \omega$. Substituting this decomposition of $\sigma$ into equation (\ref{eqn:disprel4}) and separating into real and imaginary parts gives two equations for $\gamma$ and $\omega$ as functions of $\cond$. The maximum growth rate is then found by differentiating these with respect to $\cond$, setting $\partial \gamma / \partial \cond = 0$ (so as to maximize the real part of $\sigma$),  and solving these four equations simultaneously. Here we simply state the result:\footnote{This assumes $N^2 > 0$ or, equivalently, ${\rm d} \ln T / {\rm d} \ln p < 2/5$.}
\begin{equation}\label{eqn:hbog}
\gamma_{\rm max} = \dyn \, \frac{1}{2} \left( \D{\ln p}{\ln T} - \frac{1}{5} \right) \left( \D{\ln T}{\ln p} \right)^{1/2} ,
\end{equation}
\begin{equation}
\omega_{\rm max} = \dyn \left( \frac{3}{10} - \frac{1}{4} \D{\ln p}{\ln T} - \frac{1}{100} \D{\ln T}{\ln p} \right)^{1/2} ,
\end{equation}
\begin{equation}
\omega_{\rm cond,max} = \dyn \, \frac{1}{5} \left( \D{\ln T}{\ln p} \right)^{1/2} ,
\end{equation}
so that
\begin{equation}\label{eqn:hbok}
k_{||{\rm ,max}} ( \mfp H )^{1/2} \approx 0.1 \left( \D{\ln T}{\ln p} \right)^{1/4} .
\end{equation}
This mode is overstable if
\begin{equation}\label{eqn:tcondition}
\D{\ln p}{\ln T} > \frac{1}{5} .
\end{equation}
In order for magnetic tension to significantly affect this mode, $k_{||{\rm ,max}} v_{\rm A} \gtrsim \gamma_{\rm max}$ or, equivalently,
\begin{equation}
\beta \lesssim 0.08 \, \frac{H}{\mfp} \left( \D{\ln p}{\ln T} - \frac{1}{5} \right)^{-2} \left( \D{\ln p}{\ln T} \right)^{1/2}.
\end{equation}
With typical values of $H / \mfp \sim 10$--$100$ and $\beta \sim 10^3$--$10^4$ in the outer regions of the ICM where the temperature decreases outwards, it is highly unlikely magnetic tension will affect the growth rate of this mode.

If we take into account Braginskii viscosity, it is straightforward to show that a necessary condition for stability is given by
\begin{equation}\label{eqn:hbostability}
\mc{R}' + \frac{\visc}{\cond} \frac{k^2_\perp}{k^2} + \frac{\visc}{N^2} \left( \cond + \visc \frac{k^2_\perp}{k^2} \right) > 0 ,
\end{equation}
where
\begin{equation}
\mc{R}' \equiv 1 - \left( \frac{2}{5} \D{\ln T}{\ln p} - 1 \right)^{-1} .
\end{equation}
If the temperature profile satisfies equation (\ref{eqn:tcondition}), then $\mc{R}' < 0$ and buoyant modes are potentially overstable \citep{br10}. A comparison with equation (27) of \citet{br10} reveals that Braginskii viscosity modifies this by effectively increasing $\mc{R}'$ by $\simeq$$\visc / \cond \sim 0.1$ so that equation (\ref{eqn:tcondition}) becomes
\begin{equation}
\D{\ln p}{\ln T} > \frac{1}{5} \left( 1 - \frac{1}{2} \frac{\visc}{\cond + \visc} \right)^{-1} \simeq 0.22
\end{equation}
and by further stabilising modes for which
\begin{equation}
\visc \, \cond \gtrsim -N^2 \mc{R}' ,
\end{equation}
or, using the definitions (\ref{eqn:wcond}) and (\ref{eqn:wvisc}),
\begin{equation}
k_{||} (\mfp H)^{1/2} \gtrsim 0.6 \left( \D{\ln p}{\ln T} - \frac{1}{5} \right)^{1/4} .
\end{equation}
However, the maximum growth rate of the overstability changes very little when Braginskii viscosity is included:
\begin{eqnarray}\label{eqn:newhbo}
\lefteqn
{
\gamma \simeq \gamma_{\rm max} - \dyn \, \frac{\varepsilon}{20} \left( \D{\ln T}{\ln p} \right)^{1/2}
}
\nonumber\\*&&
\mbox{} \times \left[ 1 + \frac{1}{5} \D{\ln T}{\ln p} - \frac{1}{25} \left( \D{\ln T}{\ln p} \right)^2 \right]
\end{eqnarray}
to leading order in $\varepsilon = \visc / \cond \sim 0.1$. For a fiducial cluster temperature profile of ${\rm d} \ln T / {\rm d} \ln p = 1/3$, the Braginskii term amounts to a correction $\lesssim$$16$ per cent. Braginskii viscosity does not significantly affect the fastest-growing overstable mode.

\section{Discussion}\label{sec:discussion}

The low degree of collisionality found in astrophysical plasmas such as the ICM causes heat and momentum transport to become anisotropic with respect to the magnetic field direction. This implies anisotropic heat flux and pressure. The former has been previously found to play a destabilising role in thermally stratified atmospheres, causing instabilities such as the MTI \citep[when the temperature increases in the direction of gravity;][]{balbus00,balbus01} and the HBI \citep[when the temperature decreases in the direction of gravity;][]{quataert08}, as well as $g$-mode overstabilities \citep{br10}. In this paper we have concentrated on the consequences anisotropic pressure has for the stability of the ICM.

We have argued that one cannot consider the limit of fast conduction along field lines while neglecting the Braginskii pressure anisotropy. Although there is a timescale disparity between the two effects -- anisotropic heat conduction acts on a timescale a factor $\sim$$10$ shorter than does pressure anisotropy -- both are generally much faster than (or at least as fast as) the dynamical timescale. Since the MTI and HBI occur with a growth rate comparable to the dynamical frequency, pressure anisotropy affects their dynamics significantly.

In the case of the HBI, its propensity (or, more accurately, its need) to generate fluctuations along the background magnetic field suffers from the requirement for particles in a weakly collisional plasma to conserve their first and second adiabatic invariants. The HBI changes the field strength to linear order, which induces a pressure anisotropy, which manifests itself as Braginskii viscosity and kills off the motions that generated the change in field strength in the first place. The only motions to entirely escape this constraining effect of pressure anisotropy -- those that are Alfv\'{e}nically polarised -- are also those that are stable to the HBI. The fastest growing HBI modes no longer occur at large parallel wavenumbers, but rather at wavenumbers satisfying the timescale ordering $\cond \gtrsim \dyn \gtrsim \visc$, or
\[
3 \, k_{||} ( \mfp H )^{1/2} \; \gtrsim \; 1 \; \gtrsim \; k_{||} ( \mfp H )^{1/2} .
\]
Perturbations whose wavelengths along the background magnetic field are smaller than the thermal-pressure scale-height by at least a factor $\sim$$2 \pi ( \mfp / H )^{1/2}$, while potentially unstable to the HBI, are nevertheless strongly damped. Small-wavelength perturbations whose wavevectors have a component perpendicular to both gravity and the background magnetic field behave like modified Alfv\'{e}n waves that are only slowly growing or decaying (depending on their exact wavevector orientation; see eq. \ref{eqn:alfvenhbi}). Unless $b_z \beta \lesssim H / \mfp$, Braginskii viscosity -- not magnetic tension -- sets the maximum unstable parallel wavenumber.

The situation with the MTI is more complicated. The standard MTI has a slight preference for wavevectors with projections in the $x$-$z$ plane that are aligned with the background magnetic field (see the top row of Fig. \ref{fig:mti_kxkz_angles}). This obviates heat exchange with any background heat flux, a stabilising effect when the temperature increases in the direction of gravity. Pressure anisotropy reinforces this preference, since perturbations whose projected wavevectors are not perfectly aligned with the background magnetic field are subject to strong viscous damping (see bottom row of Fig. \ref{fig:mti_kxkz_angles}). We have also found that many modes that were considered MTI-stable [e.g. $k^2_y = k^2_\perp \geq k^2 b^2_x / b^2_z$] or slowly growing become unstable in the presence of Braginskii viscosity and grow at the maximum possible rate for a given background magnetic field orientation (see Fig. \ref{fig:mti_kpky}). This is because, when $\bb{k} \bcdot ( \eb \btimes \bb{g} ) \ne 0$, Braginskii viscosity couples the Alfv\'{e}n- and slow-mode branches of the dispersion relation, so that slow-mode perturbations excite a buoyantly-unstable Alfv\'{e}nic response. By damping perturbations along magnetic field lines, pressure anisotropy frees these modes from the unfavourable consequences of having local field-line convergence/divergence.

We anticipate that many of the results found by numerical simulations of the MTI and HBI will change both quantitatively and qualitatively when the equations including both anisotropic heat and momentum transfer are implemented. This will likely have important consequences for our understanding of the thermodynamic stability of the ICM. 

Depending on the degree of collisionality in the cool cores of galaxy clusters, the field-line insulation found in many simulations to be a consequence of the non-linear evolution of the HBI \citep[e.g.][]{pqs09,brbp09} might be attenuated. This is because the large wavenumbers required to keep the HBI in action as the magnetic field becomes more and more horizontal are strongly suppressed by the pressure anisotropy they generate. Moreover, the wavenumbers at which the HBI survives largely unsuppressed have parallel components too small to rigorously be considered local, especially as the HBI reorients the mean field to be horizontal (see eq. \ref{eqn:kmax}). For a fiducial cool-core temperature profile ${\rm d} \ln T / {\rm d} \ln p = -1$, the parallel wavelength of maximum growth is equal to the thermal-pressure scale-height when $b_z \approx 110 ~ \mfp / H $. It is therefore tempting to speculate that, in the absence of strong turbulent stirring by an external agent, there exists a link between the degree of collisionality in cool cores and the mean direction of the magnetic field.

In the outer regions of non-isothermal clusters the non-linear evolution of the MTI may be more vigorous than previously thought, since many modes classified as stable or slow-growing are actually maximally unstable. Moreover, the fact that Braginskii viscosity couples damped $k_y \ne 0$ slow modes with MTI-unstable Alfv\'{e}n modes, a feature not present in the standard MTI, may profoundly affect the non-linear evolution of the magnetic field. On the other hand, the nonlinear excitation of the MTI out of its linearly-stable end state ($b_x = 0$), which is triggered by buoyantly-neutral horizontal motions \citep[see][]{mpsq10}, is unlikely to be affected by Braginskii viscosity. These motions occur perpendicular to the magnetic field (i.e. $k_\perp = 0$) and are therefore undamped by Braginskii viscosity. In either case, the spectrum of unstable modes will certainly be different, not only due to the presence of a parallel viscous cutoff but also because pressure anisotropy significantly modifies the dependence of growth rate on wavenumber.

The heat-flux--driven buoyancy overstability elucidated analytically by \citet{br10} and numerically by T. Bogdanovi\'{c} (private communication) is not significantly affected by pressure anisotropy. This is because the overstability occurs at sufficiently small wavenumbers such that conduction (and therefore viscosity) is not overwhelming (see eqs. \ref{eqn:hbok} and \ref{eqn:newhbo}). Braginskii viscosity only shifts the stability boundary slightly (eq. \ref{eqn:hbostability}).

The nonlinear evolution of the MTI and HBI in the presence of anisotropic viscosity is, at least in principle, amenable to numerical simulation. The Athena code affords one promising venue, as it is already set up for the inclusion of both anisotropic conduction and anisotropic viscosity (J. Stone, private communication). In practise, however, the implementation of pressure anisotropy into a numerical code is rather nuanced. If the pressure anisotropy exceeds $\sim$$1 / \beta$, very fast microscale instabilities (e.g. firehose, mirror) can be triggered, which will grow rapidly at the grid scale and wreak havoc upon a simulation if left unchecked. Exactly how such instabilities nonlinearly saturate remains very much an open question \citep[e.g. see][]{shqs06,sckrh08,rsrc10} and, in lieu of performing a full kinetic calculation, important choices will need to be made by the simulator regarding anisotropy limiters. Despite these rather foreboding complications, properly simulating the ICM with equations that include both anisotropic heat and momentum transfer would be a major step forward in our understanding of the dynamical stability of the ICM.

\section*{Acknowledgments}

I am indebted to Alex Schekochihin for sharing with me his expertise on multi-scale plasma astrophysics. This work has benefited greatly from his encouragement and guidance, as well as from his detailed comments on several drafts of this paper which led to a much improved presentation. I also thank Steve Balbus, Michael Barnes, James Binney, Steve Cowley, Felix Parra, and Alessandro Zocco for useful conversations. This work was initiated during the programme on `Gyrokinetics in Laboratory and Astrophysical Plasmas' at the Isaac Newton Institute for Mathematical Sciences in 2010; I am grateful to the INI for its hospitality. Material support was provided by STFC grant ST/F002505/2.

\bsp
\label{lastpage}

\end{document}